\documentclass[9.5pt,journal,final,finalsubmission,twocolumn]{IEEEtran}

\usepackage{cite}      

\usepackage{graphicx}  

\usepackage{psfrag}    

\usepackage{subfigure} 

\usepackage{url}       

\usepackage{amsmath}   
\usepackage{epsfig}
\usepackage{dcolumn}
\usepackage{bm}

\begin{document}

\title{Relativity as a Consequence of Quantum Entanglement: A Quantum Logic Gate Space Model for the Universe}


\author{Dr. John S. ~Hamel \\ Full Professor \\ Department of Electrical \& Computer Engineering \\ University of Waterloo \\ Waterloo, Ontario, Canada, N2L 3G1 \\ jhamel@uwaterloo.ca}

%

\date{April 25, 2009}

\maketitle

\begin{abstract}
Everything in the Universe is assumed to be compromised of pure reversible quantum Toffoli gates, including empty space itself.
Empty space can be configured into photon or matter gates simply by swapping logic input information with these entities through the phenomenon of quantum mechanical entanglement between photons and empty space Toffoli gates.
The essential difference between empty space, photons and matter gates are the logic input values of their respective Toffoli gates.
Empty space is characterized by an inability for the logic inputs to influence the internal logic state of its Toffoli gates since the control lines are set to logic $0$.
Photons and matter are characterized by Toffoli gates where the control lines are set to logic $1$ enabling their logic inputs to control their internal logic states allowing for their interaction according to the laws of physics associated with reality.
Photons swapping logic input information with empty space results in the propagation of light.
Photons facilitating the swapping of information between matter and empty space gates leads to the laws of motion including relativity.
This model enables the derivation of many physical laws from purely quantum mechanical considerations including the Heisenberg Uncertainty Principle, the Lorentz transformations of special relativity, and the relationship between relativistic energy and mass.
The model provides a possible explanation for many physical phenomena including dark matter, anti-matter, and an inflationary Universe.
\end{abstract}

\section{\label{sec:intro}Introduction\protect\\}

Reversible quantum logic gates are being used almost universally today as a means to understand modern physics.
Their concept enables atoms, photons, and molecules to be used as quantum computers.
They have found uses in modern theories of quantum gravity to explain Black Hole physics.

Despite the success is using this type of model to explain and exploit the laws of quantum mechanics, grand unified theories are still struggling to unify both quantum mechanics and gravity in the form of the special and general theories of relativity.
Quantum field theory (QFT) that joins group theory and quantum mechanics has been highly successful in providing a unified approach to understanding the three fields of Nature that are governed by the laws of quantum mechanics, namely the strong, weak, and electro-magnetic forces in the form of the Standard Model (e.g. \cite{weinberg:v19:67}).
Theories such as super string theory (e.g. \cite{michio:book:99}) and loop gravity (e.g. \cite{ling:v61:2000}) have attempted to include gravity by extending the Standard Model using similar approaches to QFT.
Unfortunately it has been noticed that gravity does not appear to share fundamental properties with the other three forces and attempts to unify it using the same concepts has created complications.
This could be because gravity may be fundamentally different than the other forces of Nature arising from different principles.
Although any processes that depend upon symmetries, invariance and laws of conservation can be described using field dynamics, these considerations alone are not sufficient to fully understand the nature of gravity.

In contrast to the other forces, gravity arises from the empty vacuum of space itself without requiring the presence of energy or matter for its existence.
The other forces involve particles that can actually be shown to exist in experiments.
Although energies in particle accelerators are now approaching levels where it is believed that particles associated with the weakest force will be detected, it is not clear that these particles are fundamental to the existence of gravity or are simply consequences of colliding elementary particles at high energies that enable them to form because of other fundamental aspects of empty space that are yet to be fully understood.

The success of the focus of standard approaches to understanding physics on particles with wave like properties to predict the existence of such particles does not necessarily mean that these particles are responsible for the laws of physics or if they are consequences of other more fundamental and simpler processes at work that allow such particles to exist. 
Particle accelerators and quantum field theories that predict them allow us to determine what particles are allowed to exist but do they explain why they exist as they do?

This paper will attempt to lay down a foundation of a theory of physics that does not assume the particles with their wave properties are the most fundamental or easiest way to understand the laws of physics.
Instead it will be assumed that empty space itself is comprised of basic hidden reversible quantum logic gates in the form of Toffoli gates that are connected in three dimensions into one large mesh that forms the Universe.
The gates are hidden only in that their control lines are set such that their logic inputs do not control their internal logic states.
It will further be assumed that energy quanta, in the form of elementary photons that are not necessarily the photons of the electro-magnetic force but comprise all known force exchange particles in more complex logic interconnections between Toffoli gates, are responsible for configuring empty space gates to become either these elementary photons or matter itself.
Both the most elementary forms of photons and matter are modelled as Toffoli gates identical to that of empty space but where the control lines are set such that their inputs can control their internal logic states that allows them to interact to form the laws of physics as we experience them in a reality. 
Photons differ from matter Toffoli gates in that photons are quantum mechanically entangled with empty space gates that allow them to trade logic input information including the control input information that enables the photons to propagate themselves at the speed of light.
Matter Toffoli gates are identical to photon gates but are not quantum mechanically entangled with empty space gates and therefore require the presence of photons to trade their input information with empty space to facilitate their movement.
Physical objects with rest mass that are referred to as matter arrays, are comprised of both matter and photon Toffoli gates that are quantum mechanically entangled with one another such that they can trade input and control logic information.
Matter arrays behave like ring oscillators or oscillators within oscillators possessing photons with different distributions of directional momentum that determine their motions according to the laws of relativity.
One way matter gates within matter arrays may have formed is from photon gates of opposite momentum combining to become quantum mechanically entangled with one another.

Toffoli gates are known to be universal gates for classical logical functions as well as being the simplest form of reversible quantum gate that can also be configured to be matter and energy while at the same time providing for a means to have space, and perhaps dark matter, to co-exist as entities that can influence gravity but yet not interact with other forces of nature such as electro-magnetic energy.
Other quantum gates, such as Hadamard, phase and $\pi / 8$ gates required for a Universal set of gates for a Quantum Turing Machine (QTM), can be seen as actions on Toffoli gates rather than actual structures or physical entities.
Hence it will be assumed that the universal entities underlying all structure within the Universe are configurable Toffoli gates with different logic levels at their inputs and perhaps different logic polarities associated with these inputs.
Internal random state changes within each gate limited by a maximum switching speed dependent upon the gate energy are also assumed to effect actions such as quantum mechanical entanglement interactions between gates as well as the other known possible QTM interactions that lead to known behaviour of matter and energy.
Focus will be placed on how quantum mechanical entanglement plays a role in the laws of motion.

It will be shown that this model naturally enables the theory of special relativity, i.e. the Lorentz transformation, to be derived from purely quantum mechanical considerations where quantum mechanical entanglement plays a key role.
It also enables an explanation of how the force of gravity arises in the presence of empty space logic gates of varying spatial volume that results in the bending of light in the presence of matter and energy in a multi-dimensional gate space geometry.
Using the same assumptions both the Heisenberg Uncertainty Principle can be quantitatively derived and the fact that light must travel at a constant speed can be proven from quantum mechanical theory without simply assuming that it is true from experimental observations.

The only experimental results that need to be included to make the theory quantitatively correct are 1) the measured proportionality between energy quanta and their frequency through Planck's constant $h$, 2) the measured proportionality between thermal energy and temperature through Boltzmann's constant $k$, and 3) the measured speed of light itself $c$ \cite{michelson:v34:1887}. 

This paper can be seen as the equivalent to Einstein's theory of special relativity \cite{einstein:v17:1905} but where quantum mechanical principles are used to derive the Lorentz transformations as opposed to using classical field considerations.
In other words, in contrast to existing quantum field theory, the classical fields are derived from pure quantum mechanical considerations as opposed to assuming them first and then re-expressing them in discrete form following laws of symmetry as is done in QFT.

In existing grand unified theories and the Standard Model that exploit quantum field theory, mathematical symmetries are introduced in an effort to find those that universally describe the nature of the forces and particles observed in experiment.
The use of logic circuits to explain physical phenomena is no different in that it will be shown that it is possible to converge on logic circuit systems that embody experimental behaviour.
Instead of focusing on particles, waves, strings, or loops as fundamental entities, pure information in the form of logic levels (i.e. logic $0$'s and $1$'s) are used in conjunction with symmetrical reversible logic gates to represent what is fundamental as to what is responsible for the reality we experience in Nature.
Particles then become combinations of these gates in the form of ring oscillators were all gates within the oscillator interact with one another on average once per random cycling of the internal logic states of each gate.
Instead of conventional ring oscillators familiar to electronic engineers, quantum mechanical entanglement between all gates within the matter oscillator or array of gates are assumed to be ``adjacent'' to one another such that all of them interact with another gate (or with themselves) at once in parallel simultaneously until after the same number of clock cycles pass as the number of gates within the object.
Another way to picture this is that the matter array is a digital binary register and another identical register overlaps the first moving one position relative to the original one each clock cycle so that each gate from each position in each register overlaps or interacts each shifted position.
Gates interact by swapping logic input information through quantum mechanical entanglement.
Matter arrays are assumed to be composed of pure matter gates and photon gates, both configured from empty space gate but where photon gates are the only type that can become entangled with empty space gates.
By swapping information with empty space gates, since all gate types, matter, photon and space gates, are identical of logical form, swapping information amounts to propagation or movement.
The photons must assist the matter gates in propagating by first swapping information with them and then swapping that information with space to effectively move the matter array as well.

This model of a object leads naturally to the correct formulation of the Lorentz transformations of relativity.
Only basic conceps are presented in this paper to lay the foundations of the approach.
However, in the Discussion section suggestions are made as to how to extend the model to include other physical principles such as anti-matter and charge.
What becomes apparent is that it is possible to find suitable gate elements that can be combined in hierarchical oscillators within oscillators to represent forms of matter including quarks, lepton, baryons, force exchange particles etc. that can then be built up into electrons, neutrons, protons, atoms, molecules, solids, liquids and gases, etc.
By finding suitable logic circuits incorporating more and more physical principles, one is effectively building up a unified grand theory that has the has the potential to naturally incorporate gravity and relativity into the model.
This paper does not accomplish all of these tasks but provides a model of matter, empty space, and photons that obeys the laws of gravity and relativity following already established quantum mechanical principles.
It should then be possible to modify the basic gate elements eventually moving to three dimensional gates using the same principles.
An advantage of the theory is that it is possible to derive the results of classical field theory directly from quantum mechanical principles.
Finding suitable gate structures and interactions  to develop a unified theory is no different than searching for suitable mathematical symmetries that provide invariances in the laws of physics under translation and rotation in quantum field theory.
An advantage though using the gate approach is that the theory provides explanations as to why things are the way they are from a physical vantage point that may be more understandable and more transparent provided one has an understanding of logic gates and Boolean algebra.

Section \ref{sec:gate_space} describes models for empty space, energy quanta (elementary ``photons''), and matter that consists of energy and matter simply being forms of empty space itself.
In section \ref{sec:lorentz} From this ``gate space'' model of matter and energy it is then possible to derive the laws of motion according to special relativity (i.e. the Lorentz transformations of energy, space, and time) as well as central results pertaining to quantum mechanics such as the Heisenberg Uncertainty Principle.
These results bring about certain principles that are discussed more completely in Section \ref{sec:principles}.
The assumptions of the gate space model can be used to describe a number of physical phenomena.
These phenomena are discussed qualitatively in Sections \ref{sec:inflation} to \ref{sec:stars} followed by some conclusions in Section \ref{sec:conclusions}.

\section{\label{sec:gate_space}Gate Space\protect\\}

To develop the theory the only experimental results that are required are 1) the proportionality between the energy of a quanta and its frequency being Planck's constant $h$, 2) the measured statistical proportionality between the energy of a quanta and its temperature being Boltzmann's constant $k$, and 3) and the measured constancy of the speed of a energy quanta propagating through space independent of the inertial frame of reference of an observer being the speed of light $c$. 

\begin{figure}
\begin{center}
\includegraphics{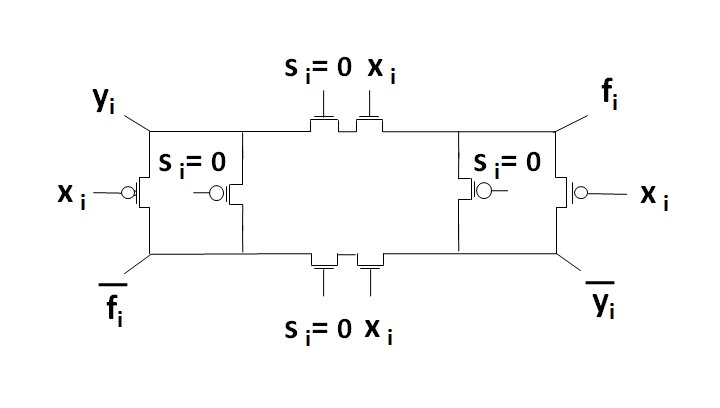}
\end{center}
\caption{Space gate modelled by a reversible Toffoli gate where $s_i = 0$}
\label{fig:sgate}
\end{figure}

\begin{figure}
\begin{center}
\includegraphics{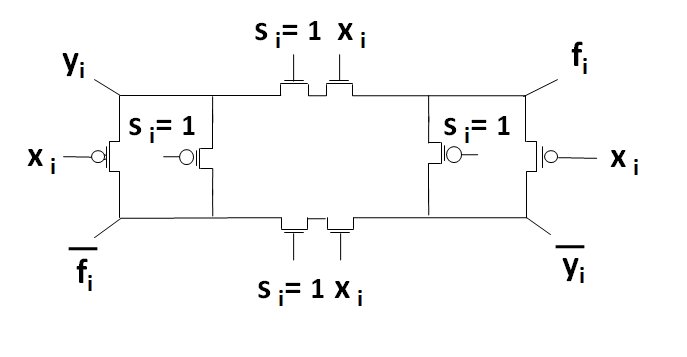}
\end{center}
\caption{Photon gate modelled by a reversible Toffoli gate where $s_i = 1$}
\label{fig:pgate}
\end{figure}

\begin{figure}
\begin{center}
\includegraphics{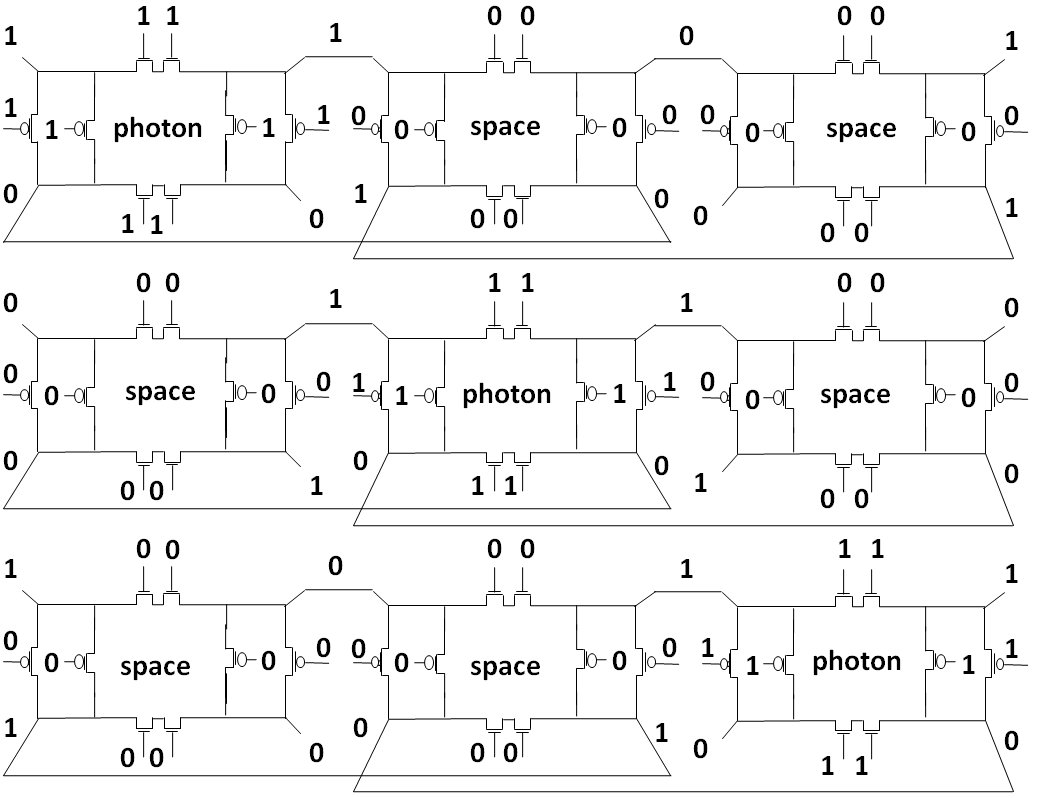}
\end{center}
\caption{One dimensional array of space gates with one photon gate propagating through space programming one space gate per $\Delta t$ to become a photon gate swapping logic values of $s_i$ control and $x_i$ input lines between photon gate and adjacent space gate by means of quantum entanglement. When a photon moves both its $s_i$ and $x_i$ information is copied in this process. Time goes forward going downward in the figure.}
\label{fig:garray}
\end{figure}

\begin{figure}
\begin{center}
\includegraphics{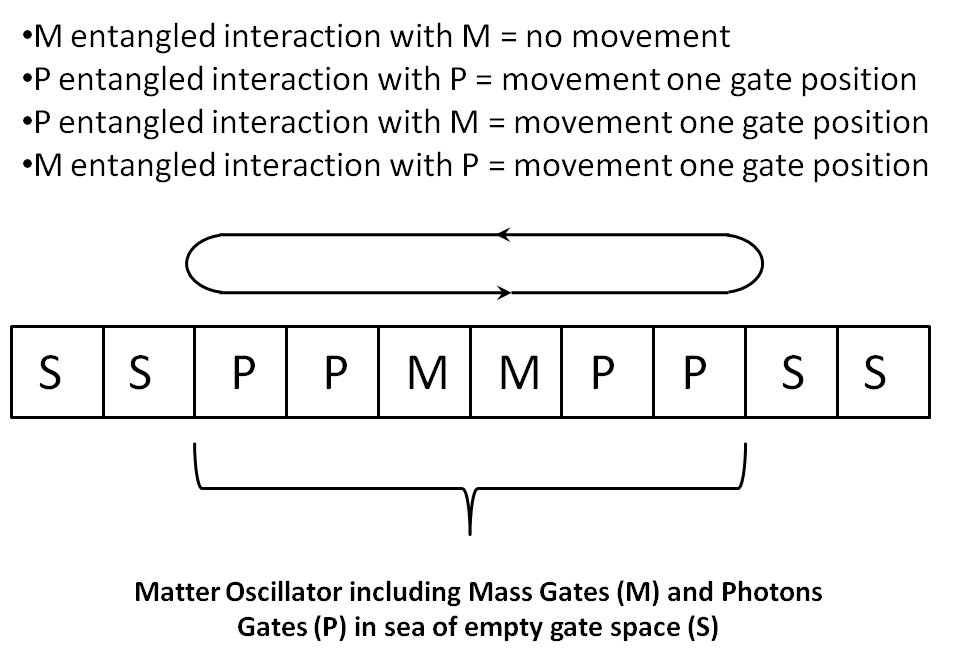}
\end{center}
\caption{Physical matter object modelled by collection of matter and photon gates in a ``matter array'' acting as a ring oscillator but where all gates are adjacent to one another through quantum entanglement regardless of complexity and dimensionality.}
\label{fig:marray}
\end{figure}

To form this theory it is only necessary to adopt the possibility that empty space itself is composed, at its most fundamental level, of two logic state reversible Toffoli gates where the inputs to the gates do not control the logic state but are cut off from being influenced by energy quanta.
This can be modelled by a gate such as in Figure \ref{fig:sgate} where the control input line $s_i$ is set to a logic zero and the other input $x_i$ is the line that interacts with an outside observer attempting to measure or influence the logic state of the gate itself $f_i$ and $\overline{f_i}$ in the circuit shown.
The circuits being used in this discussion in these figures are being depicted as CMOS logic gates where there are both NMOS and PMOS.
The NMOS transistor symbols imply that if a logic $1$ is placed on its gate then it shorts out the source and drain of the transistor creating an equi-electro-chemical potential surface in the logic line it controls.
Conversely the PMOS transistor symbol represents a transistor switch where a logic $0$ placed on its gate shorts out the source and drain of the transistor creating an equi-electro-chemical potential surface through it.

It can be further assumed that it is possible that these empty space gates are quantum entangled with one another in one form or another.
Varying degrees of entanglement might be assumed such as nearest neighbor only entanglement or all gates in the Universe being adjacent to one another.
Quantum entanglement is modelled here where one gate being entangled with another is capable of swapping control line $s_i$ and perhaps also its input $x_i$ logic values after one complete logic state cycle time $\Delta t$.
When a quanta of energy moves it is assumed that both the control line $s_i$ and input line $x_i$ information are traded with that of empty space or matter gates through this process.
If two gates are considered to be quantum entangled with one another then they are adjacent in that one can swap both the control line and input line logic values with the other in just one $\Delta t$ period.
Space gates being entangled with one another would not necessarily have any consequences swapping control line logic values since by definition all such logic values are assumed to be the same, i.e. logic $0$ in this model preventing the logic $x_i$ line, that is responsible for photon and matter gates being observable or measurable, having any impact on the empty space gate logic state $f$.
However, the possibility of empty space gates being entangled with one another even if not adjacent in connectivity via their $f_i$ and $\overline{f_i}$ outputs, can have a significant impact on how photons and matter gates might move through such a network. 

Simple analysis of this circuit shows that the functional output $f_i$ and it complement $\overline{f_i}$ forms the controlled control NOT function of a Toffoli gate.
Which input to the Toffoli gates is the control line is arbitrary but is labelled as $s_i$ that controls whether or not the logic circuit output state will respond to a change of logic signal on the other input line $x_i$.
Viewed in this manner the control line $s_i$ can be said to control the logic function of the circuit.

Quantum mechanical entanglement is then being modelled formally as the ability of one gate to control the logic function of another through the internal logic output state $f_i$ of the first gate.
If two gates are connected in this manner where one control the logic function of the other or they control each others logic functions by being interconnected in this fashion through feedback, the two gates as a system will violate the Bell Inequalities if they are viewed as being two entangled qubits.
If the gates are Toffoli gates then to accomplish this it is only necessary to arrange for the gates to control each other's control line inputs $s_i$ by their respective internal logic states $f_i$.
If the control line $s_i$ is logic $1$ for the circuits shown in Figures \ref{fig:sgate} or \ref{fig:pgate} then these circuits will have the logic function $f_i = x$ or $f_i = \overline{x}$ depending upon the value of $y_i$. 
If the control line is logic $0$ then the logic functions become constant logic functions independent of the input logic line $x_i$.
Empty space is then considered to be composed of constant interconnected Toffoli gate logic functions whose internal logic states are determined only through random internal fluctuations rippling through space.
Matter and photons that represent reality are composed of interconnected logic functions $x$ or $\overline{x}$ that form the complex things we see in the Universe since the control lines of every gate that is considered to be matter or photon energy has the property of its internal logic output state being controllable by its own logic input $x_i$.
Using this approach then enables one to model everything in existence as mere pure logic information where the logic information as logic $0$'s and $1$'s move from place to place and not the gates themselves that provide the framework for the information to interact.

A fundamental energy quanta can then be modelled by the same kind of gate as for empty space but where the control line $s_i$ is set to unity as shown in Figure \ref{fig:pgate}.
The photon gates have the additional property of being entangled with the empty space gates.
They do this by being able to swap the control line and input line logic values with an adjacent empty space gate when encountering it in a particular direction.
This direction represents the momentum or velocity direction of the photon.
In one dimension there would be only two such directions.
This directionality is a property of the photon as observed in Nature.
The manner in which the photon swaps control line and input line values with an adjacent space gate makes the photon appear to move or propagate through space resulting in its original gate becoming an empty space gate and the next gate becoming the photon gate.

Figure \ref{fig:garray} depicts arrays of these gates in a one dimensional model that could represent empty space for the gates where $s_i = 0$.
There is also a photon in the figure connected to these empty gates where the photon gate is identical to the empty space gates but where $s_i = 1$.
By having $s_i = 1$ the photon gate is becoming a real photon that can interact with other photons or matter gates through the input $x_i$ that will influence its logic state.
There is no way to observe or interact with the empty space gates when their control lines are set to $s_i = 0$ and as such represent the virtual logic states of space.

Matter gates are represented by identical gates to the empty space gates and photon gates where their control lines $s_i = 1$ as for the photon gates.
However, the matter gates are not entangled or coupled with empty space gates as photons.
As such matter gates simply stay where ever they are when they switch logic states internally.

It is further assumed that all types of logic gates, all being connected to one another as in Figure \ref{fig:garray} for a one dimensional example, have randomly fluctuating internal logic states where their internal $f_i$ and $\overline{f_i}$ values cycle through two logic states returning to the original state on average every characteristic period $\Delta t$.
The photon gates then propagate themselves advancing one gate through ``gate space'' every $\Delta t$ on average.
Figure \ref{fig:garray} depicts a photon moving through gate space in a one-dimensional model where time is progressing as one moves from one row of gates to the next going downward in the figure.

It is further assumed that all gates have a basic spatial extent or volume that will be referred to as $\Delta x$ in a one dimensional model for these discussions.
Hence the speed of light $c$ is simply the ratio of $\Delta x$ to $\Delta t$ being a constant, but where $\Delta x$ and $\Delta t$ may not be individually constant.

In some discussions a one dimensional model will be assumed to simplify explanations. 
The essential arguments, however, used to connect quantum entanglement with relativity are multi-dimensional in nature.
What is being attempted here is to construct a physics at its simplest most fundamental level upon which the actual laws of physics can be built to agree with observation.
We will see that this basic physics has the same properties as our actual physics.
The use of these reversible Toffoli gates are the simplest structures that can exist and actual structures may require several gates to realize but that interact in the fundamental ways described as per these simple fundamental structures.

It is worthwhile considering what is occurring in gate space where it is assumed that each gate has an internal random fluctuation of its logic state and how it would propagate randomly through gate space similar to what is considered to be true for virtual photons in QED.
In a sense the ``system clock'' for the Universe is being considered to be local to each gate and random  but with characteristic clock periods $\Delta t$ that depend upon the temperature of the gate as opposed to existing as some kind of Universal global synchronous clock signal keeping the Universe in sync.
When this random gate logic state process is imposed upon a network of empty space gates one can see that the influence of one gate changing will ripple throughout the space gate network.
This ripple will be stopped when encountering an opposite travelling ripple begun by a distant gate switching logic states randomly.
One can then see that there will be a random spatial and temporal distributions of these ripples of various lengths and durations.
The existence of longer ripples will have less probability of occurring or existing then short duration and short distance ripples where all travel at the speed of light.
Longer ripples will appear to have longer wavelengths.
If they travel through gates of higher energy where their size and switching speeds vary accordingly then the ripples will also have varying frequency content.

These ripples could be responsible for phenomena such as the Casimir Effect where forces exist as a result of their distributions being affected by nearby matter gates denying the existence of longer wavelength random ripples between objects that are also composed of gates but where the gates within objects would interrupt the random virtual space ripples since their logic states are controlled by their logic input lines in contrast to empty space gates.
This discussion is admittedly rather vague but it can be seen that such a model should agree with what is already known from other theories involving the random creation and annihilation of virtual particles of differing wavelengths and frequencies.

Objects with rest mass are then modelled as arrays of both matter and photon gates.
Each gate is considered to be quantum mechanically entangled at the Toffoli gate level where these gates can interact in a way where they can swap logic input and control input logic information within one characteristic clock interval $\Delta t$.
To accomplish this it is also assumed that each gate is adjacent to one another which is a single gate spatial distance $\Delta x$ due to being quantum mechanically entangled regardless of their actual real space orientation within the object. 
The object will be referred to as a ``matter array'' where both matter and photon gates are considered to be present.
The matter array is assumed to behave like an oscillator, or a hierarchy of oscillators within oscillators, where all gates interact with one another through the process of quantum mechanical entanglement swapping information with each other randomly. 
If there are $m$ matter and $p$ photons within the matter array then each gate interacts with itself or another gate on average every $\Delta t$ but all do so simultaneously such that there are $m+p$ such individual interactions per $\Delta t$.
On average every gate interacts with itself or every other gate once in this fashion in $(m+p)^2$ $\Delta t$ time intervals repeating the process over and over again in this amount of time on average.
Only photon gates $p$ can interact with empty space gates where there may be many such gates within the array or around it.
When a photon interacts with an empty space gate it swaps its information with it appearing to move to its location.
A photon can only interact with an empty space gate that is connected to it directly in real space and in a particular translational direction according to its assigned momentum direction.
If the distribution of momentum direction of photons in a matter array over all three spatial direction is uniform then the object will only appear to vibrate randomly which is called Brownian motion or thermal noise.
Matter gates cannot couple directly with empty space gates and need a photon gate to facilitate information trading between matter gates and empty space gates for the matter gates to move.
This is accomplished by the photon gates trading information with matter gates and then trading that information with empty space gates which then effectively enables a matter gate to couple with empty space gates.
For simplicity it will be assumed when a photon and matter gate interact that this results in movement of matter gate into an empty space gate position in a characteristic time $\Delta t$ when developing the laws of mechanical motion.
Applying an unbalanced external force to the matter array amounts to altering the number and also the distribution so photons with a particular momentum.
If the distribution of momentum of the photons becomes unbalanced, the number of unbalanced uncompensated photons will be referred to as $l$ as opposed to the total number of photons $p$ in the matter array.
The balance or unbalance of photons with a particular momentum direction is only relative to another matter array in an inertial frame of reference as per the normal considerations of relative motion.
For rotational motion it is possible that there are unbalances of photons in more than one perpendicular direction in the matter array leading to circular or orbiting motion.
Internal spin of elementary sub-atomic particles can be modelled by such a multidimensional unbalance of photons (e.g. as $l_x , l_y, l_z$) as an inherent property of these particles that can be related to quantum numbers.
This description of objects with rest mass that can range from elementary particles to macroscopic objects such as planets is necessary to derive the Lorentz transformations that formulate the laws of motion according to the theory of relativity.

\section{\label{sec:entanglement}A Logic Circuit Model for Quantum Entanglement\protect\\}

In order to derive the Lorentz transformations from purely quantum mechanical considerations, it is necessary to be able to model quantum entanglement from a classical reversible logic gate perspective.
Although quantum entanglement is understood phenomenologically using quantum logic circuits, more precision in understanding the phenomenon is required.

\begin{figure}
\begin{center}
\includegraphics{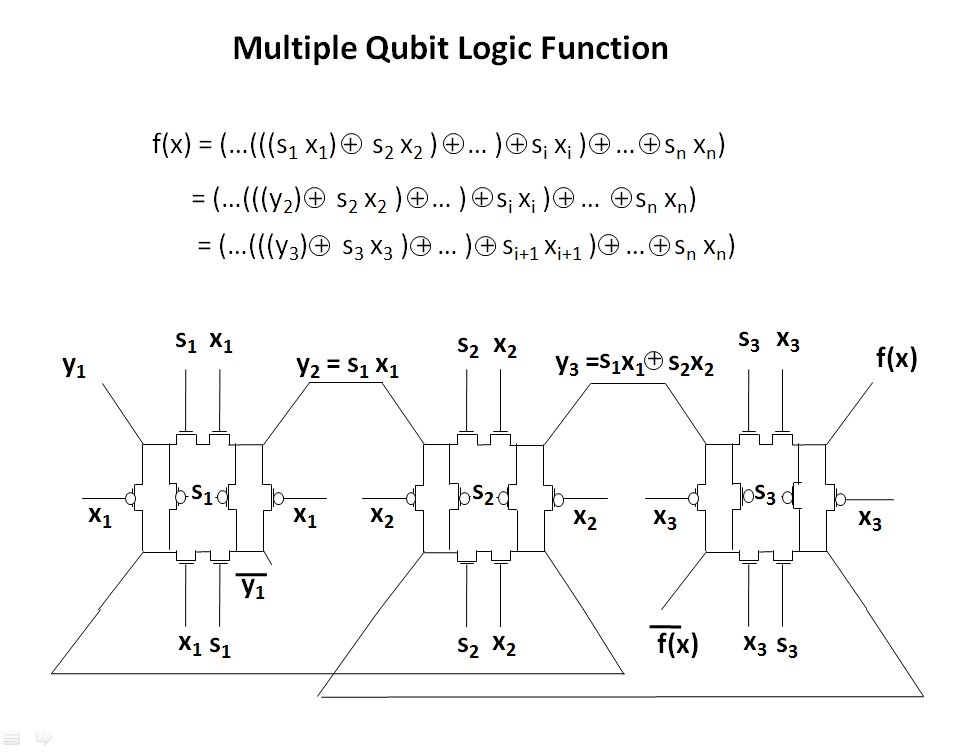}
\end{center}
\caption{1D gate space model where there is no entanglement between gates leading to a fully separable state logic array with 3 qubits in this example.}
\label{fig:ent1}
\end{figure}
\begin{figure}
\begin{center}
\includegraphics{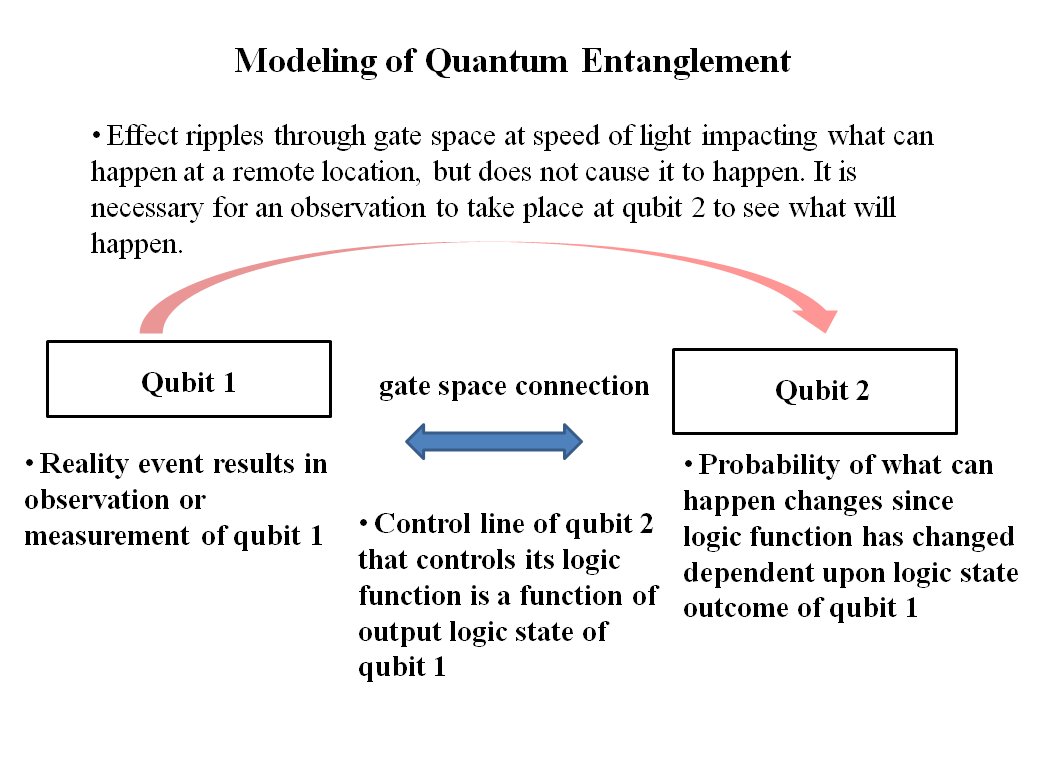}
\end{center}
\caption{Conceptual model of gate space where two qubit circuits are entangled such that its logic function and therefore its probability amplitude if measured is impacted by the logic state of another remote logic gate qubit.}
\label{fig:ent2}
\end{figure}
\begin{figure}
\begin{center}
\includegraphics{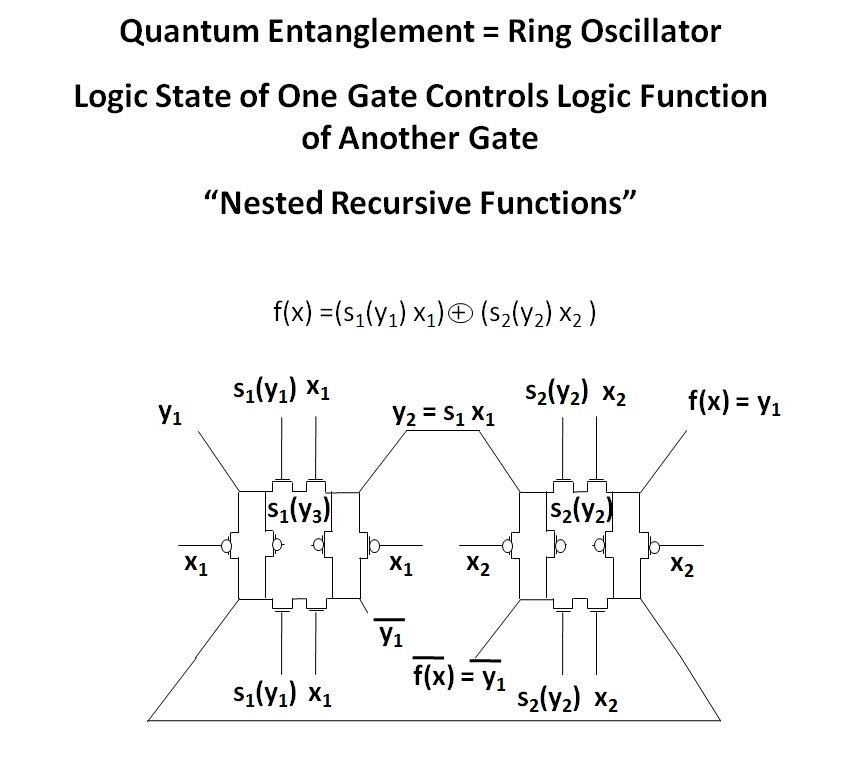}
\end{center}
\caption{Quantum mechanical entanglement can be modelled where the logic states ($f_{i-1}$ and $\overline{f_{i-1}}$) of each gate are coupled to one another's control lines $s_i$ and/or input logic lines $x_i$.}
\label{fig:ent3}
\end{figure}
\begin{figure}
\begin{center}
\includegraphics{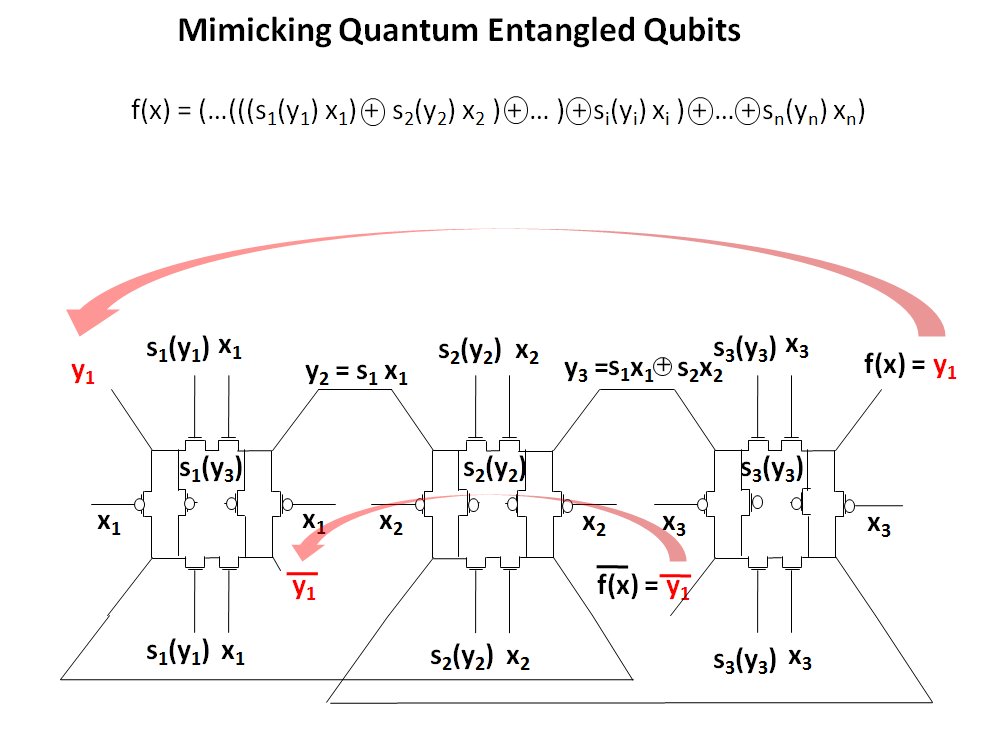}
\end{center}
\caption{Example of a 1D gate space with only 3 qubits where each gate is entangled with each other gate being wrapped around from beginning to end of the array. Such an array will be inherently unstable resulting in a ring oscillator where oscillations are sustained through continual swapping of control line and input line information between gates.}
\label{fig:ent4}
\end{figure}
\begin{figure}
\begin{center}
\includegraphics{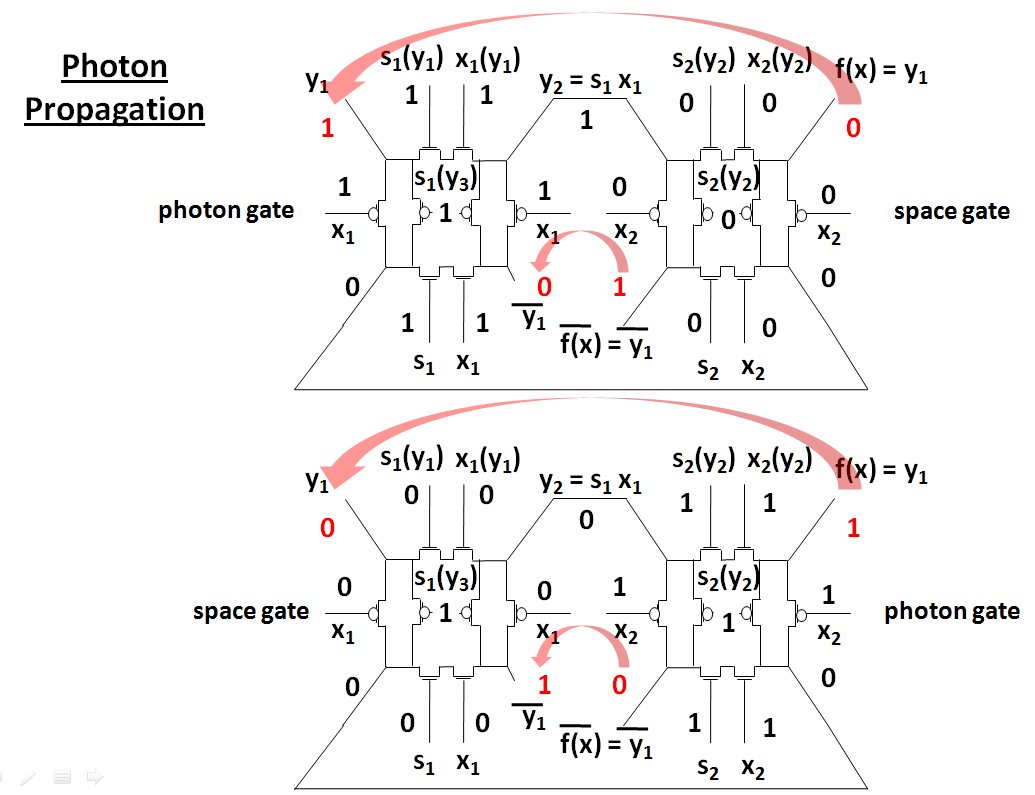}
\end{center}
\caption{Example of how a photon gate can move through gate space by swapping control and input line information with an empty space gate through the quantum entanglement model where the photon travels as if part of a ring oscillator spanning all of the Universe (only 2 qubits used to represent entire Universe in this simple example).}
\label{fig:ent5}
\end{figure}
\begin{figure}
\begin{center}
\includegraphics{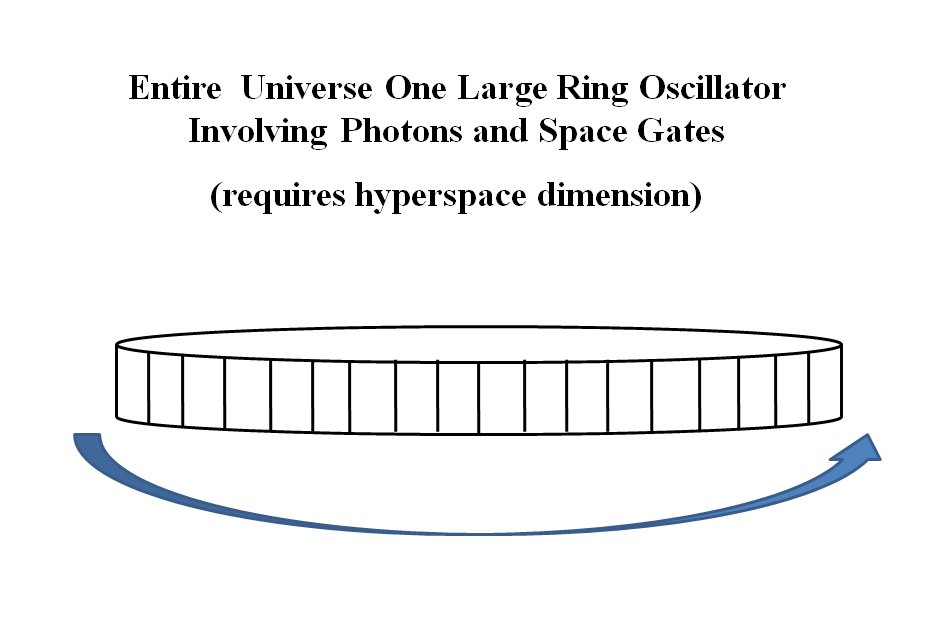}
\end{center}
\caption{A simple 1D oscillator model of the Universe where the entanglement model results in the entire Universe being composed of giant ring oscillators wrapped around on themselves in hyperspace that propagate photons simply due to the instability of the oscillators that drive them through the empty space gates.}
\label{fig:ent6}
\end{figure}
\begin{figure}
\begin{center}
\includegraphics{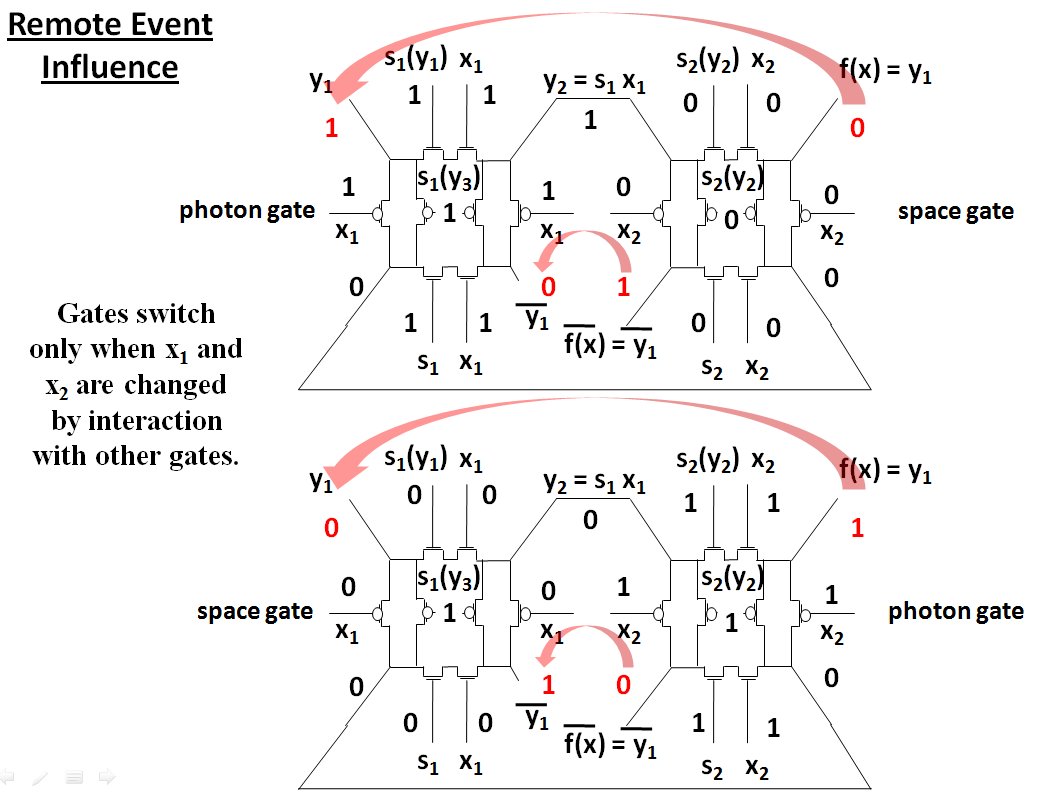}
\end{center}
\caption{Example of how two qubits are entangled but where the swapping of information is controlled dependent upon the input line logic levels $x_i$ in each gate. In this way a measurement on one gate that might result in its $x_i$ being set of logic $1$ will impact what will occur if another remote gate is measured at some other time since its logic function has been affected by the first gate through its control line $s_i$ being connected to the logic state of the first gate, even if there are many empty space gates in between the two that would transmit this information at the speed of light.} 
\label{fig:ent7}
\end{figure}
Figure \ref{fig:ent1} depicts three qubits joined together where there is no entanglement between them.
As such they form a fully separable state XOR function that is also known as an EVEN-ODD parity function generator or stabilizer circuit.

Quantum mechanical entanglement is often described as where the measurement of one qubit changes the probability amplitude of another if they are entangled. 
Figure \ref{fig:ent2} depicts one way this might be modelled using logic gates.
Quantum entanglement can be modelled using Toffoli gates if the control $s_i$ and/or input lines $x_i$ of each gate is controlled the inputs to the gate $y_i$ and $\overline{y_i}$ that are the logic levels of the function outputs $f_{i-1}$ and $\overline{f_{i-1}}$ of the other gate if they are connected together as shown in Figure \ref{fig:ent3}. 

As can be seen in Figure \ref{fig:ent4} by connecting the gates in this manner, where the inputs $s_i (y_i)$ and $x_i (y_i)$ are functions of the $y_i$ input being the output $f_{i-1}$ of the previous gate, that the array becomes inherently unstable resulting in a ring oscillator.
These oscillators being depicted in these Figures are in real space, but if all gates within a matter array are considered to be adjacent to one another being quantum mechanically entangled with one another, then the ``ring'' oscillator becomes a hyperspace ring oscillator.
There are many ways in which this coupling can be modelled using classical circuits that would be equivalent, where the ones shown are only intended to provide a means to crudely model a more sophisticated interconnection that would occur in multi-dimensions in real matter and energy.

What is important to notice, however, is that this model of quantum entanglement leads to a oscillating behaviour.
Matter and energy are often considered to be composed of fundamental harmonic oscillators in standard models of physics, and as such, the approach being used here is consistent with this widely accepted view.

A consequence of quantum entanglement leading to oscillating behaviour is that the movement of photons through space, as depicted in Figures \ref{fig:ent5}, itself can be viewed as ring oscillators encompassing the entire expanse of the Universe wrapping around on itself in a higher dimensional hyperspace as depicted in Figure \ref{fig:ent6}.
It is these oscillations that provide the ``clock'' that drives the logic gates throughout space.

Matter arrays form local ``clocks'' in that their oscillations remain local.
Photons in matter arrays, however, are also coupled to the Universal oscillator structure of space resulting in either Brownian motion if their momentum distribution is balanced in any particular inertial frame of reference, or resulting in movement in any other frame of reference if there is an imbalance of photon momentum directionality within the array.
Matter gates might be viewed as photon gates of opposite momentum direction that have become entangled with one another forming a local oscillator thereby appearing to be matter gates in a particular frame of reference, or perhaps there exist pure matter gates that do not couple with space.
Either can be assumed when deriving the Lorentz contractions with equal effect.

Matter arrays then also influence the remainder of gate space including empty space gates through their photons coupling with empty space.
In this way there are oscillators within oscillators where eventually the highest hierarchy oscillator is the Universe itself in a particular momentum direction.
Matter being composed of elementary particles leading up to atoms and then to molecules may involve a large number of elemental oscillating matter arrays that perhaps make up the elementary particles themselves at a deeper levels within the particles, or perhaps there is a particle already known to exist that is a pure single matter array.
Regardless of the level within matter where it is organized as matter arrays, they act in parallel according to the Lorentz transformations as will be derived in a later section, when an object moves mechanically in any particular direction or combinations of directions in non-linear movement.
The result does not depend upon how matter is organized into these fundamental oscillators, just that it does.
How these fundamental oscillators behave controls all of physics regarding the interaction of matter and energy quanta including how the object exists in time.

Empty space itself can also have such matter arrays but that are dark not interacting with photons if the inputs to the Toffoli gates are always at complementary logic levels that prevent any one logic input line from controlling the logic state of the gate.
This arrangement will prevent normal matter and energy from interacting with the empty space matter array (or dark matter) but the array will still result in gravitational attraction and can still produce oscillations that might influence other phenomena, such as the Casimir effect.
These dark particles can have any number of gates or length (wavelength) and oscillation periods (frequency spectra) that are not detectable through electro-magnetic means.
Dark matter will be discussed in more detail in a later section.

Figure \ref{fig:ent7} depicts how one logic gate can influence the probability of events that might take place at a remote location in gate space in another logic gate or array of logic gates.
If the output of a gate or gates $f_{i}$ and $\overline{f_i}$ that represents their internal logic states, are connected to other gates by influencing their control lines $s_i$, then a measurement of the first gate or gates (representing an observable event that took place in reality) can influence what might take place at a remote location thereby influencing the probability amplitude of the remote gate or array of gates since they are connected by empty space gates.
This influence only takes place in the simple example shown in Figure \ref{fig:ent7} when the logic lines of both gates involved are set to logic $1$ in the example.
The $x_i$ logic line in the first gate represents an event taking place there which then influences the control line of the remote gate.
Then, at some time later, when an event that takes place causing the $x_i$ input line to go to logic $1$ in the second gate, what logic levels $f_i$ that gate can take has been influenced by what took place in the first gate.
By influencing the logic function itself of a remote gate, the probability of what internal logic levels this gate can possess has been altered. 
If entanglement exists even just locally between empty space gates adjacent only in real space, as opposed to within matter arrays where it is being assumed that all gates are mutually entangled with one another directly, the impact of gates within a matter array being measured or observed can influence any remote matter array gates through this process since both matter arrays are entangled with empty space through their photon gates.

In essence, all matter within the Universe has this potential influence where any event can influence what might happen anywhere else.
In macroscopic situations this impact might not be noticeable since these influences do not cause remote events to be connected directly, only their probabilities of occurring.
Perhaps the summation of all of these connections between events (measurements or observations that influence gate inputs) and probabilities is responsible for statistical averages involving the law of large numbers etc.
However, when extremely miniature systems are studied at the quantum level, the degree of freedom can be restricted to the extent where the phenomenon of quantum entanglement can been directly detected where there is a hundred percent correlation between measurement and probability outcome between two qubits as would occur in the simple circuit models of the Figures presented here.
In macroscopic systems it is not possible to eliminate all of the potential influences on the probability amplitudes of another system being observed except through observing a large number of events that leads to the laws of statistics we observe in Nature.

\section{\label{sec:lorentz}Derivation of the Relativistic Lorentz Contractions as a Consequence of Quantum Entanglement\protect\\}

In this section the Heisenberg Uncertainty Principle of quantum mechanics, the Lorentz transformations or contractions according to the theory of special relativity, and the relationship between relativistic energy and mass according to the theory of General Relativity, are all derived assuming the same set of quantum mechanical principles according to the logical properties of Toffoli gates being used to model empty space, photons and matter. 
No use is made of results derived from classical or quantum field theory.
The fields themselves can be derived from the assumption of empty space being composed of Toffoli gates where they can be configured by photons to be matter or photons.

Quantum mechanics is embodied by one key experimental result, that of the proportionality between energy quanta and frequency through Planck's constant $h$.
Relativity is embodied by the key experimental result of the constancy of the speed $c$ of propagation of these energy quanta in the form of photons.
Finally, the essential concepts of thermodynamics rests upon the key observation that the energy of these quanta is proportional to their temperature through Boltzmann's constant $k$.
Using these experiments and an appropriate assumption of the connectivity nature of matter, energy and empty space itself, it is possible to formulate a foundation upon which the whole of physics can be constructed.

The goal here is to provide this foundation by going back to the beginning of modern physics where these experimental results were first noticed.
Before one can proceed to unifying physics, which is well beyond this paper, it is first necessary to combine these observation in a unified manner that can then be utilized to develop a complete unified theory including all known forces and interactions between energy, matter and free space.
As such it is first necessary to be able to quantitatively derive the basic foundations upon which quantum mechanics, special relativity and general relativity are based.
These include 1) the Heisenberg Uncertainty Principle, 2) the Lorentz transformations of special relativity, and 3) the relationship between energy and mass including the nature of gravity.

This paper will attempt to derive 1) and 2) including part of 3).
It should be possible to derive the gravitational force of attraction between masses from the model presented here but this is beyond the scope of this paper.
The gate space model presented here naturally leads to a geometrical model of how the spatial volume of gates varies with energy and mass in multi-dimensions.
Using this fact as presented it should be possible to derive the laws of gravity using the same methods already well established from differential geometry and non-Euclidean mathematics.

It will first be shown that the mere assumption that empty space is comprised of reversible Toffoli gates of finite spatial extent and finite switching speed, and that can be configured by photons to be either photons themselves or matter leads to the Heisenberg Uncertainty Principle and the fact that there must a maximum speed of any entity comprised of these gates propagating through space regardless of the mass of the object, its energy, or its frame of reference.
These assumptions also lead to the well known relationship between energy and mass according to the general theory of relativity.
It will then be shown that by assuming that there exists quantum entanglement between photon and matter gates within an object with a rest mass (a matter array) where photons are responsible for the propagation of both photon and matter gate through empty space, that the Lorentz transformations of special relativity result.

By equating the Boltzmann distribution of energy within any physical system of total energy $E$ at a temperature $T$ that is composed of any number $n$ of Toffoli gates that each contain two logic states one has, 
\begin{eqnarray}
\exp \left( \frac{E}{kT} \right) = 2^n
\label{eq:expE}
\end{eqnarray}
from which we obtain 
\begin{eqnarray}
E = nkT \ln{2} 
\label{eq:En}
\end{eqnarray}
where $n$ is the total number of gates in the physical system of energy $E$ be they matter or photon gates.
Only gates where the control line $s_i = 1$ are considered to have observable and measurable energy.
This is because any physical observer using any kind of measurement or sensory system, be it natural or artificially constructed, are also composed of matter array gates comprising matter and photon gates.
The measurement system connects to the system being measured via the input lines $x_i$ of each individual gate in the system through complex arrangements that may also include being entangled quantum mechanically.
Any information flowing into the measurement system does so through changes in the inputs $x_i$ of the system being measured influencing the logic states of the measurement system through its own inputs $x_i$.
This is why it is not possible to measure systems composed of interconnected gates where the control lines $s_i$ are set to zero.

It is interesting to consider, as a minor diversion, the possibility of complex objects existing in gate space that may be identical in form or logic construction to real objects we can observe but where the control lines of every gate in the virtual system has been set to logic $0$ cutting off any interaction with the physical world as we know it.
This model provides the possibility that there may exist such complex objects of interconnected space gates within our own Universe with which we cannot interact.
One does not have to assume that such virtual gates only form simple connections as depicted in Figure \ref{fig:garray} representing empty space.

Setting $n$ to 1 for one gate we obtain the well known result that a single bit of information contained in one of these gates is given by,
\begin{eqnarray}
E = kT \ln{2} 
\label{eq:E}
\end{eqnarray}
We know experimentally that the energy contained within a quanta is,
\begin{eqnarray}
E = h f
\label{eq:Eh}
\end{eqnarray}
where $h$ is Planck's constant and $f$ is the frequency in $Hz$.
For a single two state logic gate its state changing frequency can be related to its characteristic state clock period $\Delta t$ such that,
\begin{eqnarray}
f = \frac{1}{2 \pi \Delta t}
\label{eq:f}
\end{eqnarray}
We then equate the minimum energy $E$ of one of the logic gates at a temperature $T$ to its Planck energy such that,
\begin{eqnarray}
E = h f = \frac{h}{2 \pi} \frac{1}{\Delta t}
\label{eq:Egate}
\end{eqnarray}
from which we obtain the relation for the two logic state gate,
\begin{eqnarray}
\Delta E \Delta t = \frac{h}{2 \pi} = \hbar
\label{eq:Hgate}
\end{eqnarray}
Assuming the total energy contained within the single logic gate is equally divided between the two logic state according to an equi-partition concept, one obtains the Heisenberg Uncertainty Principle for the minimum energy detectable $\Delta E$ as,
\begin{eqnarray}
\Delta E \Delta t \geq \frac{1}{2} \frac{h}{2 \pi} = \frac{\hbar}{2}
\label{eq:HE}
\end{eqnarray}
Thus, assuming that quanta are essentially binary logic gates at their simplest level with characteristic logic state changing times one can derive the Heisenberg Uncertainty Principle directly from the Planck proportionality between energy and frequency.
Naturally, if everything is made from these kinds of gates then there can exist no measurement instrument that can resolve an energy less than that of one of logic states within one of the gates with any degree of certainty, hence the limit to the measurement of the two parameters of energy and time.

The momentum can be derived through the fundamental definition of a force $F$ being the rate of change of momentum $\Delta p$ with time $\Delta t$ for a single gate.
The momentum can then be derived from,
\begin{eqnarray}
\Delta E = F \Delta x = \frac{\Delta p}{\Delta t} \Delta x
\label{eq:E3}
\end{eqnarray}
where we obtain the expression,
\begin{eqnarray}
\Delta E \Delta t  = \Delta p \Delta x \geq \frac{1}{2} \frac{h}{2 \pi} = \frac{\hbar}{2}
\label{eq:HE2}
\end{eqnarray}
giving rise to the alternate form of Heisenberg's Uncertainty Principle.
From this we obtain for the minimum momentum of a single state within a logic gate that has two states,
\begin{eqnarray}
\Delta p = \frac{\hbar}{2} \frac{1}{\Delta x}
\label{eq:p1}
\end{eqnarray}
where we see the momentum is inversely proportional to the gate spatial extent.
The momentum associated with the gate, however, it taken as twice this amount since there are two logic states and therefore two ways to store energy.
This is a matter of interpretation as to what minimum energy or momentum is measurable.
In standard existing models of physics energy quanta are traditionally modelled by invisible cavities in empty space that contain $n + 1/2$ $\hbar \omega$ energy where one half wavelength within the cavity is the longest, lowest energy state of the photon or energy quanta.
The gate model amounts to the same assumption but where there is an invisible two logic state circuit in place of the cavity and energy is no longer in the form of a wave.
An oscillator of the form assumed by traditional physics could not exist in such a simple form as it takes many photons to give wavelength behavior in experiments such as the double slit experiment.
A single photon would only create a dot on the screen beyond the slits where many photons are needed to statistically build up a wave like pattern.
As such a photon does not have true wavelength.
If one assumes that the energy is pure information with the connectivity behaviour of a logic circuit then the lowest order of that energy, if one assumes that the energy is equally divided between the two logic states, $E = (h/2) f$ = $(1/2)\hbar \omega$ as per the harmonic wavelike oscillator model of traditional physics.
Instead of concentrating on a single logic state, the minimum energy usually referred to will be that of an entire two state logic circuit being $E = h f = \hbar \omega$ and for $n$ such gates $E = n \hbar \omega = nkT\ln{2}$ where we will ignore the $1/2$ term.
If we insist on keeping the $1/2$ term to agree with traditional physics we will always end up including an extra half logic gate that can be allowed but seems awkward.
Another approach, if one wanted to use precisely the same conventions as traditional physics, is to just assume that the energy of a gate is $E = (1/2)kT\ln{2} = (1/2) h f$ but this might not agree with Planck's experimental results.
As such there appears to be an essential difference between assuming energy quanta are waves in cavities or are contained within logic circuits by this small factor.
It will be assumed here that the logic circuit approach is more correct since it includes quantum mechanical entanglement that is not possible in a cavity in the normal sense other than to simply impose that property on the microwave cavity.
Quantum mechanical entanglement will be defined in terms of what takes place between logic gates that is natural and quite simple to understand.

For a single quanta of energy or matter comprised of only one gate, using the two forms of the Heisenberg Uncertainty Principle we obtain,
\begin{eqnarray}
\frac{\Delta E \Delta t}{\Delta p \Delta x} =  1
\label{eq:Hratio}
\end{eqnarray}
from which one can obtain the relation,
\begin{eqnarray}
\frac{\Delta E}{\Delta p} = \frac{\Delta x}{\Delta t} = c
\label{eq:Hc}
\end{eqnarray}
From this relation one sees that there must be a maximum speed of propagation for any influence through the empty space gates since there must be non-zero uncertainties in energy and momentum due to the discrete yet finite nature of the logic gates that comprise Nature with regards to both spatial extent and switching state changing speed.
Also it can be easily proven that the speed of propagation of energy quanta must be constant since both the uncertainties of energy and momentum scale linearly with the number of gates in the same physical system these quantities represent.
Hence, regardless of the energy of the quanta or its order $n$, the ratio of the energy to momentum uncertainty must be the same.
The energy of the quanta is related to the inverse of its logic switching period and its momentum inversely proportional to its spatial extent.
Hence the ratio of the energy to momentum uncertainties for any single quanta ($n = 1$) or number of quanta together in a physical system ($n > 1$), will be both constant and related to the speed of light.

As such the speed of propagation of quanta through the Universe naturally arises from the assumption that everything including empty space can be modelled by reversible quantum logic gates.
As such, assuming such a model for the Universe is the quantum analog of Einstein's classical assumption of of the constancy of light based on experimental observation.
Einstein also assumed that the speed of light was independent of the inertial frame of reference of the observer from any viewpoint which is also a natural direct consequence of assuming a ``gate space'' model of space itself.
Any observer in gate space is also compromised of those same gates configured to be matter and energy whose propagation through space is also limited by the same limitations as for the quanta of light, but where there are matter gates that must be moved by the photons in the observer that slows down the observer relative to light.
This results in the propagation of pure photon energy quanta being independent of any observer or physical source once launched into the gate space network.

It can be seen that this model leads naturally to a theory of gravity and relativity.
The switching speed of a single gate increases as the energy of the quanta increases according to Planck's relation since the number of logic states remains constant at two per quanta.
To maintain a constant speed of light this means that the gates that comprise a higher energy quanta, which could also represent mass, must decrease in size maintaining the ratio of its spatial size to its switching time constant.
This is actually a very physical mechanical concept in that a gate half the size of another gate might be able to change states twice as fast if its velocity of whatever is changing within were to remain constant.
In other words, if the gate were seen as a kind of linear one dimensional mass moving back and forth along a line within a box, decreasing its length allows it to decrease its period of oscillation increasing its fundamental frequency of oscillation but maintaining a constant velocity of travel as it moves back and forth along its linear one dimensional cavity.

\begin{figure}
\begin{center}
\includegraphics{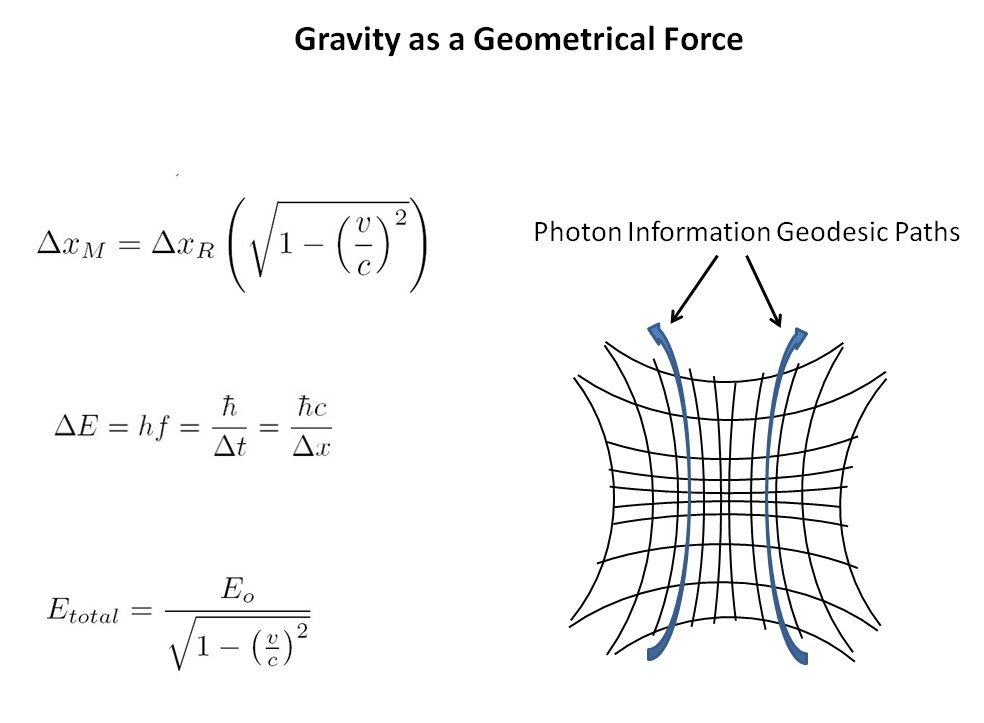}
\end{center}
\caption{Picture depicting how higher energy logic gates that have smaller spatial dimensions can in effect curve space-time through a geometrical gravitational force in multi-dimensions where photon gates will appear to be attracted to the higher energy regions (that could be matter or energy) appearing to curve through space.}
\label{fig:curved_space}
\end{figure}

From the relationship between energy $E$, switching time $\Delta t$, and spatial size $\Delta x$ of a logic gate for a single gate representing a quanta of energy $E = kT \ln{2} = h f = h / {2 \pi \Delta t} = (hc)/{2 \pi \Delta x}$,  one sees that as the energy (and momentum) increases so does its frequency that in turn decreases its switching time requiring its spatial extent $\Delta x$ to decrease to maintain a constant maximum speed of propagation $c$.
Since the speed of light is constant, so is the ratio of the quanta of energy $E = \Delta E$ and momentum quanta $p = \Delta p$ is a single gate.
This also applies to any physical system with $n$ gates since both energy and momentum scale with the number of gates in the same manner.

Therefore the higher energy quanta, which also represents mass in the form of matter arrays, will have logic gates with smaller physical size.
Assuming that a photon gate, being light, will follow a line of gates a particular direction, in a two or three dimensional space the light will appear to curve towards the smaller dimensional gates where there is more relativistic energy and mass.
This leads to a concept of gravity where the photons will tend to be attracted to the higher energy density locations in gate space.
If matter and photons travel together then the photons will lead the matter along the same geodesic.
This is depicted in Figure \ref{fig:curved_space}.
Using this knowledge in conjunction with the manner in which gates change their physical size depending upon their energy it should be possible to derive a non-Euclidean formulation of the laws of gravity using the gate space model to arrive at the same conclusions as is done now in conventional classical general relativity field theory.
How the gates change with size can be done using a thermal approach where the formulas already given above relate gate size to energy, temperature and number of gates, or it can be done through assuming the equivalence of forces associated with inertial energy according to the Lorentz transformations and the force of gravity in the presence of mass.
As already mentioned, the goal of this paper is not to redevelop the known classical laws of non-Euclidean gravity, but to lay the framework for the nature of gate space itself and how some fundamental laws of physics naturally arise from this assumption.

It should be noted, once again, that these geometries exist whether or not the control lines of the gates allow the inputs $x_i$ to the gates to influence the internal gate logic states $f_i$. 
In this way it is possible for empty space as well as arrays of empty space gates with identical arrangements to matter arrays to exist also influencing gravitational forces without being influenced by energy quanta such as photons thereby making these empty space gate arrays potentially dark matter or energy.
In fact, this appears to be an unavoidable consequence of assuming that empty space itself is composed of logic gates that pop into reality by being configured through Toffoli control line concepts.
There should be some kind of influence from these empty space gates even if their control line logic levels cut them off from interacting with normal matter and energy through the input logic lines.
If there are more empty space gates with control lines set to logic $0$ than normal matter and energy in the Universe, then to some extent they might be considered to be more important in shaping the nature of space itself compared to normal matter.
There is a symmetry between empty space gates where the control lines are set to logic $0$ in the assumed models and normal matter and energy where they are set to logic $1$ enabling normal matter-energy interactions through the input logic lines.
It is not unlike the symmetry between a photograph and its negative.
Both contain the same information but in reverse.
Perhaps we as human observers composed of normal matter and energy see the Universe from one vantage point where it might be equally valid to see it from the other perspective.
The other perspective, however, does not obey the same laws of physics if the objects within it do not interact through their logic input lines $x_i$ as for normal matter and energy.
Instead, they may interact only as pure probabilistic objects influencing each other through $y_i$ and $f_i$ inputs and outputs as opposed to the $s_i$ and $x_i$ inputs and outputs of normal matter and energy.

We will now derive the Lorentz transformations that enable a quantitative determination as to how the gates change size with energy.
This energy can come from either thermal energy of the gates switching faster due to being smaller or due to an object traveling through space being escorted by an imbalance of photons of a particular momentum in a particular direction.
Thermodynamically it does not matter whether or not the energy comes about because of a higher temperature where the photons in a matter array have a balanced momentum direction distribution leading only to Brownian motion with no net movement through space, or it has an unbalanced momentum distribution of photons as a result of an external force having been applied to accelerate the matter array to a translational or rotational velocity that too has thermal energy but directed in one direction.
As such there is a thermal equivalent between heat and inertial motion with regards to the resulting spatial size of the gates comprising the entity decreasing with total energy.
The forces involved are identical regardless of whether or not the gate size reduction leading to attraction of photons and matter arrays is due to being within a mass or being accelerated through empty space.
As such a matter array representing a large energy composed of photons and matter gates would result in the same spatial gate distortion as would the same total relativistic energy associated with the translational acceleration of a smaller matter array increasing its total energy in this fashion.

In this model only photon gates can program empty space gates to effect movement.
The photons can also program matter gates to move it through space, but matter gates cannot program empty space gates to effect movement.
A matter object comprised of solely matter gates would not move nor would it vibrate with Brownian motion.
However, there are always some photons in real matter and hence there is always movement.

\begin{figure}
\begin{center}
\includegraphics{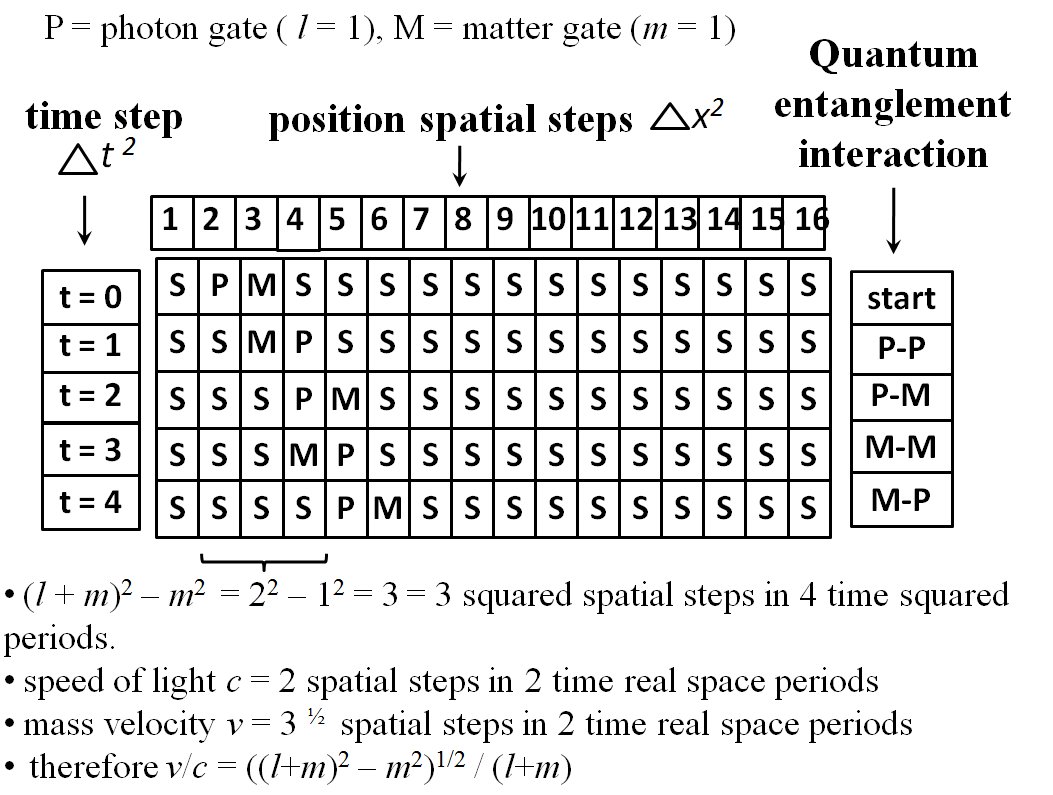}
\end{center}
\caption{Example of an object moving through space according to the laws of motion where there is one net photon $l=1$ with momentum direction in the direction of motion and one matter gate $m=1$ that cannot couple directly with space gates itself.}
\label{fig:lorentz}
\end{figure}

\begin{figure}
\begin{center}
\includegraphics{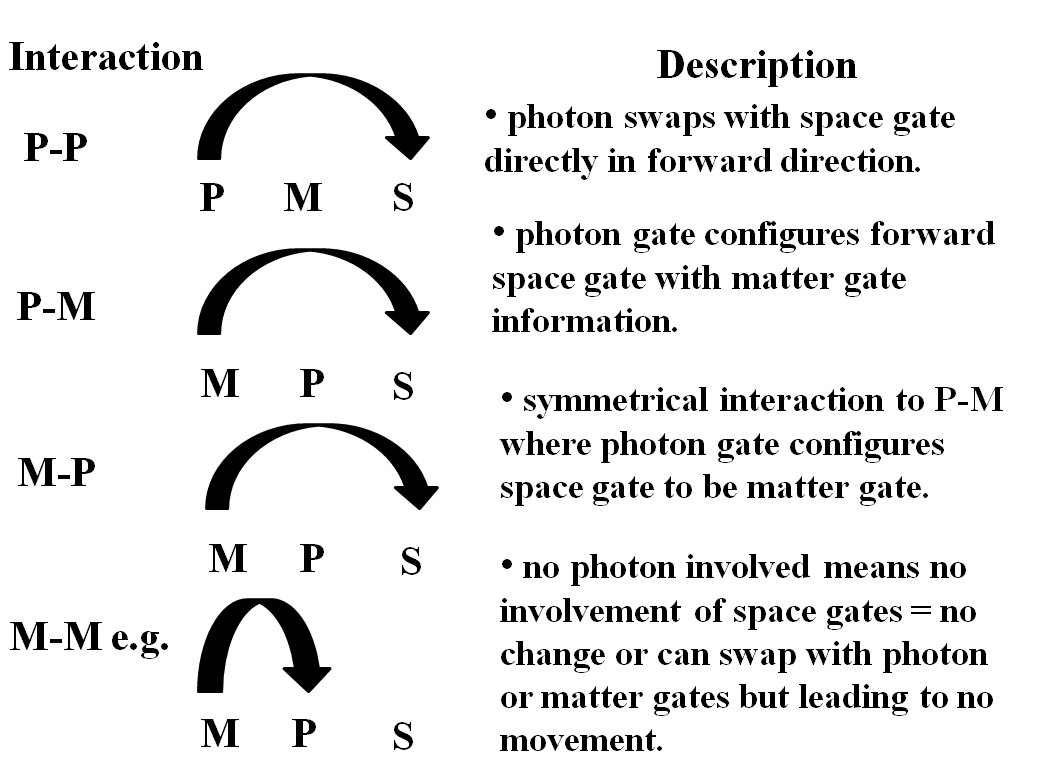}
\end{center}
S \caption{Types of quantum mechanical interactions between the various types of gates, photon gates $P$, matter gates $M$, and space gates $S$ through quantum entanglement. Quantum entanglement between photon and matter gates enables photons to copy matter gate information into an empty space gate location whereas a photon gate can copy its own information directly into an empty space gate by swapping control and input line logic levels with the empty space gate. This simple model leads naturally to the Lorentz transformations of special relativity where it is necessary to assume that all gates within an object are quantum entangled with one another at the most basic level of information representation in the Universe, and where photons are the only gates that can interact directly with empty space gates.}  
\label{fig:interactions}
\end{figure}

\begin{figure}
\begin{center}
\includegraphics{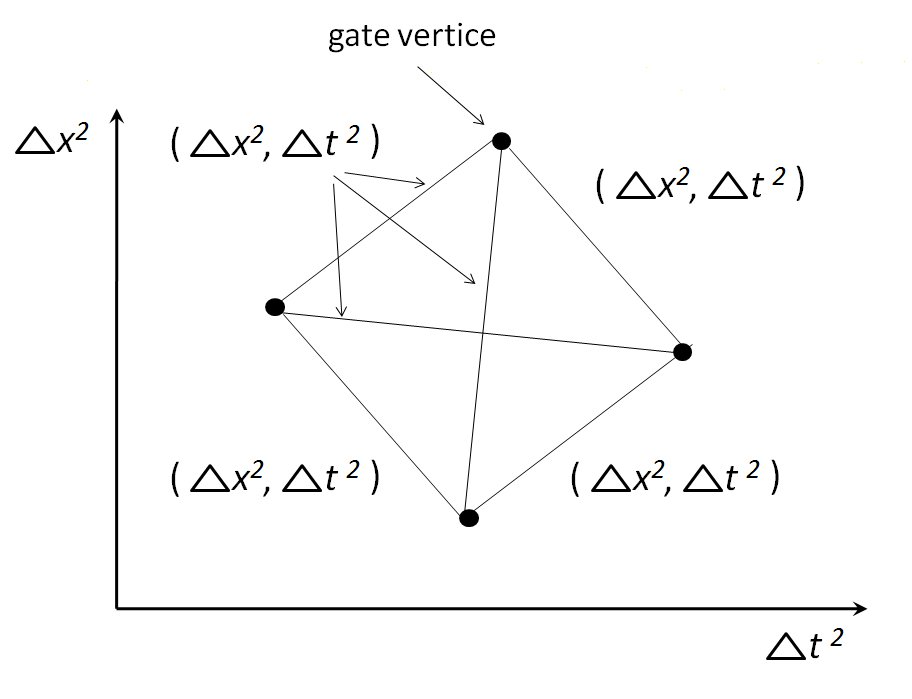}
\end{center}
\caption{Squared space-time gate continuum to model extra dimension required for one-dimensional real space-time example of an object moving through space-time. Each vertice and edge represents a gate and its interaction with another gate such that each gate and its interaction are separated by $\Delta x^2$ and $\Delta t^2$ in space and time in the new co-ordinate system.}
\label{fig:vertice}
\end{figure}

Consider a one dimensional matter system containing some photon gates and some matter gates.
Let the number of photon gates be $l$ and the number of matter gates by $m$.
An object with rest mass composed of matter and photon gates with a particular momentum direction distribution imposed by an external force to establish the photon number and momentum balance leading to translational movement through empty space has already been described in the previous section.
It is assumed that all gates within a matter array (also referred to as an object) are quantum mechanically entangled with one another in that they are able to swap control and input line information.
When such a matter array containing $m$ matter gates and $l$ unbalanced photons in a particular momentum direction, in real space-time there are $(m+l)$ gate interactions at the same time.
After $(m+l)$ $\Delta t$ intervals, due to quantum mechanical entanglement, each gate interacts either with itself or every other gate in the array.
A photon gate interacting with itself simply swaps its information with a forward empty space gate in the direction of movement.
A photon gate interacting with another photon gate does the same for the other photon gate and vice versa in a reverse interaction.
A photon gate interacting with a matter gate moves its information into an empty space gate in the direction of motion.
A matter gate interacting with a photon gate also has its own information moved to an empty space gate in the direction of motion due to the assistance of the photon.
However, a matter gate interacting with itself or another matter gate does not effect movement through space as matter gates are assumed to not have the property of being able to entangle with empty space gates to swap information.

Pure photon interactions are labelled $P-P$, photon-matter gate interactions $P-M$ or $M-P$, and matter-matter gate interactions are labelled $M-M$.
If all the gates in the matter array were photons (which means it would not be a matter array), then all photons interacting with themselves or each other, it does matter, results in one movement through space for all of the photons simultaneously where there are $(m+l)$ interactions per $\Delta t$ time interval resulting in movement of the entire array one $\Delta x$ gate forward in real space-time.
Then the entire array would move at the speed of light moving one $\Delta x$ every $\Delta t$ in real space-time.
For a matter gate where the $m$ gates are indeed matter gates and not photon gates then the entire array will move less than a $\Delta x$ in a $\Delta t$ in real space-time having a velocity $v$ less than the speed of light $c$ since the $M-M$ interactions lead to no net spatial movement of at least a portion of the matter array when they occur.
Figure \ref{fig:interactions} depicts the various types of quantum mechanical interactions that can take place between the various gates.

In real space-time it is difficult to calculate this distance for an arbitrary number of $m$ and $l$ gate arrays even in one dimension unless one uses a higher dimensional method in a new metric.
This can be accomplished by realizing that all gates in the matter array are actually adjacent to one another in both space and time through the phenomenon of quantum entanglement.
This means that it is possible for any gate to swap information or interact with any other gate in one $\Delta t$ interval in real space-time as if it is only a $\Delta x$ away in space regardless of its actual position in the array.
In reality, it may be necessary for a gate to involve other gates between it and another gate to impact its logic state.
It may also be necessary to swap gate positions while the array is moving to maintain its structural order without advancing through empty space.
These effects in real space-time increase the path length or the delay time for the matter array to move through space swapping its gate information with empty space compared to the movement of pure arrays of photons.
These increases in path length or delay time result from the necessity for photons to facilitate the movement of information from matter gates to empty space gates.

Instead of dealing with all of this complexity, a higher ordered space-time can be utilized where each real dimension for space and time are turned into two dimensional spaces each so that each gate can exist adjacent to one another being only a $\Delta x$ and a $\Delta t$ apart in space and time for a one dimensional real space-time model.
In other words, it is possible to draw a separate two-dimensional co-ordinate system for each the $\Delta x$ and $\Delta t$ intervals between each gate and each gate interaction event where the two co-ordinate systems would be a $\Delta x$ versus $\Delta x$ and a $\Delta t$ versus $\Delta t$ axis.
Combining two graphs into one we can draw a squared space-time co-ordinate system as in Figure \ref{fig:vertice} where the matter array is depicted as a graph where each vertice is a gate plus its interaction with another gate and can be a distance of $\Delta t^2$ and $\Delta x^2$ from each other gate and its interaction with that gate along edges between them with measures of $\Delta x^2$ and $\Delta t^2$ within the new vector space.
This will be called the squared space-time continuum.
Then, instead of having gates moving themselves or other gates through interactions into empty space gates advancing one $\Delta x$ of real space per $\Delta t$, one has them moving $\Delta x^2$ in each $\Delta t^2$ for photon-photon, photon-matter, matter-photon interactions.

In reality, in real space there are $(m+l)$ simultaneous interactions each $\Delta t$ interval where each gate interacts with itself and every other gate.
After a total of $(m+l)$ $\Delta t$ intervals each gate has interacted with itself and every other gate where each gate is limited to one interaction per $\Delta t$.
As such there are a total of $(m+l)^2$ gate interactions in $(m+l)$ $\Delta t$ intervals in real space-time with $(m+l)$ actions occurring per $\Delta t$ interval simultaneously.

This can then be viewed equivalently as $(m+l)^2$ individual separate sequential gate interactions each taking place in $\Delta t^2$ intervals where interactions that can move the array will advance it $\Delta x^2$ in distance in the squared space-time continuum.
Since $\Delta t$ and $\Delta x$ are both small fractions of seconds and centimeters, the square of these values are even smaller.
This does not mean that there are observable or detectable actions taking place under the Heisenberg Uncertainty Principle limit of $\Delta t$.
It means that $(m+l)^2$ individual actions separated each by $\Delta t^2$ time intervals in squared space-time are actually $(m+l)$ simultaneous actions taking place every $\Delta t$ intervals $(m+l)$ times so that all gates will interact with one another and themselves over and over again.
The new squared space-time simply allows us to exploit the inherent adjacency between all gates in both time and space through quantum entanglement.
Using this fact we are simply constructing a new spatial co-ordinate system where they actually are adjacent mathematically.
We then assume $(m+l)^2$ serial closest neighbour single gate-to-gate interactions every $\Delta t^2$ advancing through squared space-time $\Delta x^2$ each interaction so that we preserve the actual $(m+l)$ simultaneous interactions taking place advancing the matter array $\Delta x$ each time in normal space-time. 
The two space-time continuums are normalized by the fact that the speed of light is the same for both if the matter array were composed of pure photons gates.
To convert between one space-time continuum to the other one merely takes the square root of whatever is derived in the squared space-time system to go to the normal one.

The reason we are assuming that each gate will interact with itself and every other gate in this time is to take into account that in quantum mechanics, all possible path actions are known to take place to arrive at what is finally observed as a law of physics according to quantum field theory of the standard model.
This contrasts to an optimum path being taken according to classical Lagrangian physics.
As such we are assuming that a matter array or an object moving through space does so by exhibiting two known behaviours assumed of in any quantum system, that of quantum entanglement between all entities and all possible interactions taking place where their total actions are summed.
The interactions of all gates with one another and with themselves does so at $(m+l)$ interactions simultaneously while each gate finds another gate with which to interact, including the possibility of interacting with itself.
However, each gate is assumed to couple with only one other gate or itself at a time and as such there must be $(m+l)$ separate such simultaneous actions each separated by $\Delta t$ for a total of $(m+l)$ $\Delta t$ intervals before repeating interactions.
These are quite general assumptions about the behaviour of a matter array as a kind of quantum ring oscillator where all gates interact or overlap with one another and themselves to effect the overall behaviour of the object or matter array under the action of an external force.

Now that we have developed a higher order space-time continuum it becomes a simple matter to calculate how far the matter array will move in $(m+l)$ $\Delta t$ intervals in real space that corresponds to $(m+l)^2$ $\Delta t^2$ intervals in squared space-time. 
There are a total of $(m+l)^2$ separate individual interactions of the gates each $\Delta t^2$ seconds apart in square space-time each having the potential to move the entire array $\Delta x^2$ spaces.
However, we know that there are $m^2$ matter-matter interactions that do not result in spatial movement.
Hence, in squared space-time after $(m+l)^2$ $\Delta t^2$ intervals the matter array only moves $(m+l)^2 - m^2$ $\Delta x^2$ spatial intervals.

To convert back to normal space-time it is only necessary to take the square root to obtain the result that the matter array will move $\sqrt{(m+l)^2 - m^2}$ $\Delta x$ spatial gate intervals in $(m+l)$ $\Delta t$ time intervals. 
Comparing to how far light can travel (i.e. $(m+l)$ $\Delta x$) in normal space-time in that time interval the ratio of the velocity $v$ of the matter array to the speed of light $c$ in normal space-time becomes,
\begin{eqnarray}
\frac{v}{c} = \frac{\sqrt{(m+l)^2 - m^2}}{m+l}
\label{eq:vc1}
\end{eqnarray}

Figure \ref{fig:lorentz} depicts a simple example of the above for $l=1$ and $m=1$.
It can be seen that in four interactions there is one interaction of the matter gate with itself that does not lead ot movement through squared space-time.
As such the matter array only moves 3 $\Delta x^2$ spatial intervals in 4 $\Delta t2$ time intervals.
The ratio of the velocity $v$ of the matter array to the speed of light $c$ becomes $\sqrt{2^2 - 1^1} / 2$ where the matter array would be moving at the speed of light if the two gates $m$ and $l$ were to each move two $\Delta x$ in normal space in two $\Delta t$ time intervals.

The justification for using the squared space time comes about from the adjacency afforded by quantum mechanical entanglement in both space and time between all gates.
The complete overlap between gate interactions with every gate interacting with every other gate stems from the knowledge that in quantum mechanics, the laws of physics including their invariance under transformations, such as the Lorentz transformation, result from integrating or summing over all possible interactions between elementary entities, which in this case happens to be the swapping of information between Toffoli gates in gate space.
Otherwise it would not be possible to arrive at an invariant since a different set of interactions, if one did not include them all, would lead to a different expression for the transformation and hence to laws of physics depending upon the likelihood of certain interactions taking place which is not reasonable.
Also, by using a square space-time continuum exploiting the adjacency of each gate in the matter array it was possible to avoid computing probabilities involving more than two gates as would be required in normal space since in reality all gates are not adjacent to one another.
This would then require a weighted path integral method where edges to the graph would have weightings making the computation much more complex as is traditional quantum electrodynamics.
However, by creating a higher ordered space-time representation every edge of the graph has the same weighting.
Hence, there should be an analog between this approach and using QED to derive the Lorentz transformations using traditional path weighting approaches involving more than two particles including time reversal.
Time reversal is also avoided using this square space-time approach but might be involved if normal space were used throughout the derivation where photons might propagate backwards in time to maintain the order in the arrangement of gates in the matter array as it propagated through normal space.

Using the graph theory approach depicted in Figure \ref{fig:vertice} the interactions between the gates can be considered as a kind of Hamiltonian in square space-time acting on the entire matter array of gates in the object that results in movement through real space.
A Hamiltonian cycle is performed where each vertice is visited exactly once allowing it to interact with each other vertice along all edges or paths as per an Euler cycle.
In other words, the action of the gates within the matter array perform an Euler cycle visiting all paths between vertices for each vertice visited exactly once in the overall Hamiltonian cycle in the higher dimensional space-time as a result of the influence of quantum entanglement.
The laws of motion then become an information exchanging process between the photon and matter gates with empty space gates in a particular translational and/or rotational direction depending upon the photon momentum vector distribution within the object where photons are the only types of gates that can implement or facilitate the information exchange.

In this process the pure matter gate interactions $M-M$ in a physical object increases the path length causing additional delays of the object moving through space-time compared to that of pure photons.
Originally Lorentz transformations were derived, not of course by Einstein, but by Lorentz to compute how light travelled through a liquid in a tube of flowing water compared to a tube with a vacuum.
The photon scattering within the water increases the path length of the photons thereby resulting in the delay compared to photons in a vacuum.
The raw velocity of the photons themselves is not affected by passing through transparent objects.
As such the Lorentz contractions of special relativity can be interpreted as a discrete path length increase or time delay increase of electro-magnetic energy in the form of photons involving fundamental quantum mechanical interactions within matter compared to within empty space itself according to the same scattering principles.
Within matter arrays, photons, being entangled with matter gates, are delayed compared to their counterparts in empty space, by having to facilitate the movement of matter gate information into empty space gate positions. 
To make the Lorentz transformation invariant it was necessary to include all possible interactions or {\em paths} to accomplishing this in matter array in the calculation of the path length increase compared to that of photons navigating their way through empty space.

The relativistic energy in accordance with the Lorentz transformations can now be determined in agreement with the known laws of relativity from the gate space model as follows.
The rest energy of the matter array before adding photons would be,
\begin{eqnarray}
E_o = mkT \ln{2}
\label{eq:Eo}
\end{eqnarray}
and would increase linearly as $l$ photons are added becoming,
\begin{eqnarray}
E = (m+l)kT \ln{2}
\label{eq:Et}
\end{eqnarray}
where it is assumed that both the mass and energy as proportional to the number of gates in the physical system of the object.
Here $m$ can include photons whose momentum direction cancel out that can also be included in the derivation of the Lorentz transformation being modelled as gates that want to move the object in the opposite direction during some of the interactions.
These were simply modelled as matter gates that did not have the ability to trade information with empty space gates where two photons of opposite momentum direction would represent $m=2$ with the same results.

The ratio of the total relativistic energy representing the sum of the rest energy and the kinetic energy to the rest energy is,
\begin{eqnarray}
\frac{E_{total}}{E_o} = \frac{m+l}{m}
\label{eq:Er1}
\end{eqnarray}
Manipulating the ratio of the velocity $v$ to the speed of light $c$ gives,
\begin{eqnarray}
{\left( \frac{v}{c} \right)}^2 = \frac{(m+l)^2 - m^2}{(m+l)^2} = 1 - \frac{m^2}{(m+l)^2} 
\label{eq:vc2}
\end{eqnarray}
Therefore it can be shown that,
\begin{eqnarray}
\sqrt{1 - {\left( \frac{v}{c} \right)}^2} = \frac{m}{(m+l)}
\label{eq:vc3}
\end{eqnarray}
where we arrive at the Lorentz transformation for relativistic energy,
\begin{eqnarray}
E_{total} = \frac{E_o}{\sqrt{1 - {\left( \frac{v}{c} \right)}^2}}
\label{eq:Er2}
\end{eqnarray}

It is now possible to deduce the Lorentz contractions for time and space of the travelling object from the point of view of the observer at rest which is the ``gate space'' perspective.
What is in equation (\ref{eq:Er2}) can be thought as being observable from the gate space perspective which is the rest observer's frame of reference with $v = 0$.
According to this perspective, the energy of the object moving at $v$ relative to an observer at rest can be thought equivalently as an energy quanta with a higher frequency and therefore a smaller $\Delta t$.
Substituting $E = hf = \hbar / \Delta t$ on each side of equation (\ref{eq:Er2}) we obtain for the $\Delta t_R$ for the observer at rest that sees a smaller clock compared to that of the traveller,
\begin{eqnarray}
\frac{\hbar}{\Delta t_R} = \frac{\hbar}{\Delta t_M \sqrt{1 - {\left( \frac{v}{c} \right)}^2}}
\label{eq:dtr}
\end{eqnarray}
This implies that the clock for the traveller as seen by the rest observer appears to slow down as measured in the rest frame according to,
\begin{eqnarray}
\Delta t_M = \frac{\Delta t_R}{\sqrt{1 - {\left( \frac{v}{c} \right)}^2}}
\label{eq:dtm}
\end{eqnarray}
where $\Delta t_M$ is the clock period of the moving traveller as seen by and measured in the frame of reference of the observer at rest whose clock period is $\Delta t_R$ for reference.

Another way to derive the time dilation between moving and rest frames of reference is to consider how many gates exist in the elemental matter array oscillators in the moving frame of reference compared to those of the rest frame.
The moving frame has gained more energy and monentum in absolute terms where extra photons with a particular momentum distribution leading to relative motion have been added to each elemental matter array oscillator to effect this movement through gate space.
How long these matter arrays take to complete one complete oscillation where each gate swaps information with itself and with each other gate such that all gates have interacted with all other gates governs the fundamental physical interactions between matter and energy for the chemical and other physical processes in any particular frame of reference.
These fundamental oscillators are in essence the fundamental ``clocks'' that govern time itself in each frame of reference.
Yet, the laws of physics, including the speed of light, has not changed in either frame of reference from the vantage points of each observer in their given frame of reference.
The laws of physics only appear to change in the opposite frame of reference to each observer.
As such, for both observers, regardless of their relative motions and regardless of the differences in the number of photon gates within their identical matter arrays that might exist in either frame of reference, the matter arrays must complete a full oscillation cycle within $m$ $\Delta t_M$ periods in the moving frame of reference and $m$ $\Delta t_R$ periods in the rest frame of reference where $\Delta t$ is the time period between two gate interactions within a matter array or the logic cycling period of a single logic gate in either frame of reference.
However, in the moving frame of reference, in $m$ $\Delta t_M$ time periods there are $m+l$ gate interactions implying that the $\Delta t_M$ must be longer than $\Delta t_R$ by a factor of $(m+l)/m$.  
As such, from the point of view of the observer at rest, the clocks in the moving frame of reference are running slower where, once again we have, 
\begin{eqnarray}
\Delta t_M = \frac{\Delta t_R}{\left( \sqrt{1 - {\left( \frac{v}{c} \right)}^2} \right)}
\label{eq:dtm2}
\end{eqnarray}

One might ordinarily think that the gates in the moving frame of reference are speeding up as they move faster, but in this model we are modeling an increase in kinetic energy as being due to adding extra photo logic gates not increasing the switching speed of each gate.
As such the switching speed of each gate will be expected to remain the same requiring a longer time for an oscillator to cycle through all of its gates in the moving frame of reference.
Thus, it would appear to the moving observer that their physical clocks measuring time being composed of logic gates have slowed down, where the actual $\Delta t$ of each gate has not changed compared to that of the rest observer.
As such the $\Delta t_M$ and $\Delta t_R$ are not the $\Delta t$ periods of the logic gates.
For reference we can set $\Delta t_R = \Delta t$ of the logic gates in both frames of reference, but $\Delta t_M$ is longer than $\Delta t$ of each of the gates in the moving frame by a factor of $(m+l)/m$ as already explained.

To determine the spatial dilation, knowing the time dilation, the standard argument can be used that since the clocks appear to have slowed in the moving frame from the point of view of the rest frame of reference, to keep the speed of light the same in the moving frame from the perspective of the rest frame, the rest frame needs to see a smaller dimension in the direction of motion in the moving frame to compensate for the slower clock such that,
\begin{eqnarray}
\Delta x_M = \Delta x_R \left( \sqrt{1 - {\left( \frac{v}{c} \right)}^2} \right)
\label{eq:dxm}
\end{eqnarray}
where $\Delta x_M$ is the dimension of the moving traveller as seen by the observer at rest as measured in the rest frame and $\Delta x_R$ is the dimensions in the frame of rest. 

It is important to point out, once again, since photon gates are being added to the matter arrays in the moving frame of reference, from the perspective of the moving observer in their own frame of reference, neither the $\Delta x$ nor $\Delta t$ of their logic gates have changed compared to what the rest observer would observe of their own matter array gates.
However, from the point of view of an outside observer in a different frame of reference, the spatial extent of the logic gates of a remote matter array at a higher energy will appear to be smaller, regardless of whether or not the remote matter array is moving or not.

This is the equivalence of inertial forces and gravitational forces first suggested by Einstein.
As such, higher total energy matter arrays appear to have smaller logic gate dimensions from the point of view of a lower energy system thereby distorting space as depicted in Figure \ref{fig:curved_space} simply due to their being more gates and hence more energy and momentum associated with them.
Hence, it does not matter if the extra gates happen to increase energy resulting in movement or simply increase total energy as in a pure mass, they will distort space and time relative to another matter array of a different energy.
This results in forces of attraction between them in a geometrical manner.
This is constitent with the fact that time is slower on a larger massive object compared to outside its gravitational influence according to the theory of relativity.
From a quantum mechanical perspective, this distortion in space and time results from the matter arrays becoming larger due to more gates being added to them resulting in longer cycling times.

The famous relationship between energy and mass can also be derived quite simply from assuming this simple gate space model.
Energy as work is related to force per gate propagation as,
\begin{eqnarray}
W = E = F \Delta x
\label{eq:W}
\end{eqnarray}
From Newton's third law that defines what mass means we obtain,
\begin{eqnarray}
F = m_o a = m_o \frac{\Delta x}{\Delta t \Delta t} = \frac{E}{\Delta x}
\label{eq:F}
\end{eqnarray}
for each space gate that a rest mass $m_o$ propagates through.
We then obtain for energy $E$ of a mass $m_o$ of gates,
\begin{eqnarray}
E = m_o \frac{\Delta x \Delta x}{\Delta t \Delta t} = m_o {\left( \frac{\Delta x}{\Delta t} \right)}^2 = m_o c^2
\label{eq:mc2}
\end{eqnarray}
As such, using basic definitions of work, force and mass one obtains the relativistic energy $E = m_o c^2$ which is consistent with known classical relativity field theory.

In deriving the Lorentz transformation it was assumed that each of the $(m+l)$ gates in a matter array on average interacted with itself or another gate every $\Delta t$ in a simultaneous manner but that each individual gate interacted with itself or another gate sequentially every $\Delta t$ as opposed to doing so simultaneously in parallel in some manner.
This requirement for a particular gate to interact with either itself or only one other gate every $\Delta t$ can be thought of as a kind of ``principle of non-simultaneity''. 
We know that we arrived at the correct formulation for the Lorentz transform by making this assumption and indeed it is necessary for this assumption to be true to maintain a constant speed of light.

Consider what might happen if for some reason there existed two different situations for a given matter array or object.
At one time it obeys this principle of non-simultaneity where there is a $\Delta t$ delay between interactions between two given gates, but at another time a gate decides to interact with more than one gate simultaneously.
It would be possible to detect these two situations through measurement by measuring the time difference it took for the object to move from one place to another where one time it obeyed one law of special relativity and another time a different one.
Besides this, it would be possible to ascertain that two things occurred in the second instance in a vanishingly small time interval approaching zero $\Delta t$ compared to the first situation which violates the Heisenberg Uncertainty Principle.
The Heisenberg Uncertainty Principle places limits one how accurately two events can be ascertained to have occurred simultaneously with an error of at least $\Delta t$ depending upon $\Delta E$.

Several known phenomena related to relativity naturally stem from this model of gate space.
It is known that the speed of light is independent of the velocity of its source.
From the gate space model this naturally arises since the speed of light is being limited by the ratio of the spatial extent of the logic gates comprising empty space $\Delta x$ to their characteristic logic state cycling period $\Delta t$. 
The velocity of the object itself is being mitigated by the photons quantum mechanically entangled with it in its matter array that move through space by its own photons programming empty space for themselves as well as the matter gates that cannot couple with empty space.
If the source were to emit light then this would entail some of its photons that were originally entangled with the matter array becoming de-entangled and allowed to propagate on their own at the speed of light without having to waste time intervals providing motion for the matter gates that cannot couple with space themselves.

It can be seen from the presentation of this new model that the classical field theory of special relativity naturally arises from pure quantum mechanical considerations which may be in contrast to other existing theories of quantum gravity.
Other theories tend to quantize the classical field theory where the Lorentz transformations are first derived from the classical field theory originally formulated by Albert Einstein. 
General Relativity, such as the bending of light around higher energy regions of space, naturally arises from this new model since in multi-dimensions it can easily be seen that the smaller gates corresponding to the higher temperature or higher energy quanta and the space they occupy will result in a bending or distortion of the entire gate space network appearing as a depression around such a higher relativistic energy objects.
Photons will be forced to propagate along a geodesic from one gate to another that will tend to bend towards these smaller gate regions in a two or three dimensional logic gate space network.
Gravity from this perspective then becomes a geometric force due to the restrictions placed on photon movement, that also move the matter arrays that must follow their paths, by the multidimensional space gate network where the spatial extent of these gates is inversely proportional to their energy from very simple quantum mechanical considerations as derived above.

Finally, it is not surprising, once a gate space model was adopted, arriving at the correct formulation for the Lorentz transformations requires an interaction between all gates, including with themselves, sequentially as opposed to assuming some degree of simultaneity that might violate the Heisenberg Uncertainty Principle.
The interaction between all gates is in agreement with a Hamiltonian cycle concepts that are known to be present in all other theories of physics where energy operators are present.
These facts increase the reliability and credibility of the result as well as the assumed gate space model itself that is seen to preserve the correct behaviour of motion according to special relativity.

As already mentioned, the momentum in gate space is modelled as the inverse of the spatial gate extent by $p = \hbar/(2\pi \Delta x)$.
As the energy and momentum of a quanta increases its spatial extent decreases.
The entire space gate network, whether it be in empty space, within energy arrays or within matter arrays, must conform to an interconnected logic circuit. 
As such, any matter or energy propagating itself to interact with other energy or matter arrays must conform properly to the correct spatial size as well as being synchronized in time.
``Fitting'' spatially into the network conforming to the existing gate spatial extent of what is already in the vicinity amounts to the conservation of energy but in a geometrical manner.
Logic state switching speed defined by $\Delta t$ being synchronized to be compatible between interacting gate arrays within the larger Universal gate space network corresponds to energy conservation.
Other conservation principles such as compatible spin etc. can also be identified in these terms if more than one dimension is considered where energy and matter arrays can have photons with momentum direction distributions that would result in ``spinning'' gate arrays where gates are being copied into empty space gates in circular motions that might represent inherent angular momentum in elementary particles of matter. 
Spins in elementary particles might then appear more as circular rippling oscillators with different directions of rotation that can represent circularly polarized photons of light etc.

In a three dimensional model of the entire Universe, it is most likely that the gates are connected in a way that each gate is connected to other gates with no dangling inputs or outputs.
The only way this can be true is if the entire gate network were existing in a higher order dimension being at least four dimensional expanding as the temperature cool adding more and more gates according to $E = nkT\ln{2}$ to maintain a constant total energy of the Universe.
This is also what is believed to be true for the Universe in other models in mainstream field theory according to the laws established by Einstein's theory of relativity.
The concept of space being compromised of interconnected gates that are all connected together with no dangling inputs or outputs (i.e. there are no edges to the network) lead intuitively to this hyperspace model without requiring any other proof.
Either the gate network of the Universe has edges, which is not very likely, or it is continuous in three dimensions, a more likely situation, where a hyperspace is automatically required to accommodate this second more likely situation.

\begin{figure}
\begin{center}
\includegraphics{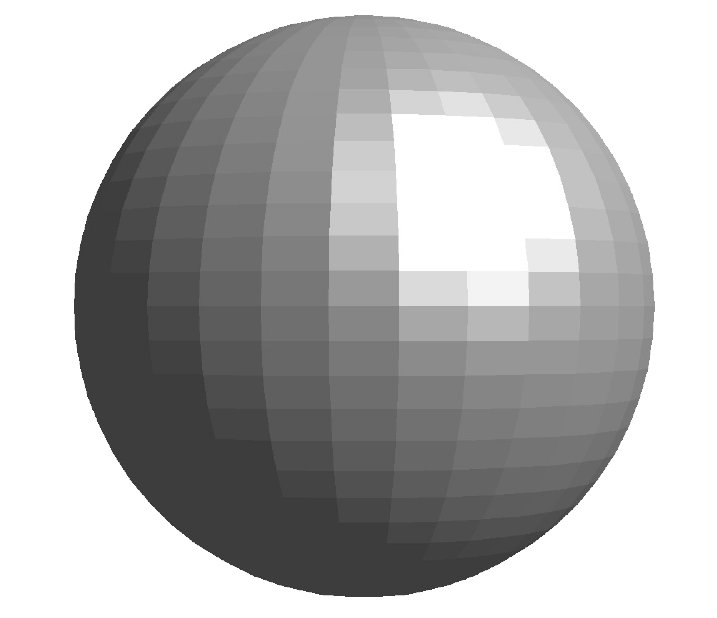}
\end{center}
\caption{Two dimensional model of the Universe composed of two dimensional gate space network that is fully connected without edges to the network. The real Universe is a three dimensional sphere or another kind of closed three dimensional ``surface'' in a four dimensional space as often assumed in the Standard Model of physics for General Relativity. The concept of a gate space network with no edges or unconnected gate inputs and outputs naturally leads to this same model of the Universe.}
\label{fig:sphere}
\end{figure}

\begin{figure}
\begin{center}
\includegraphics{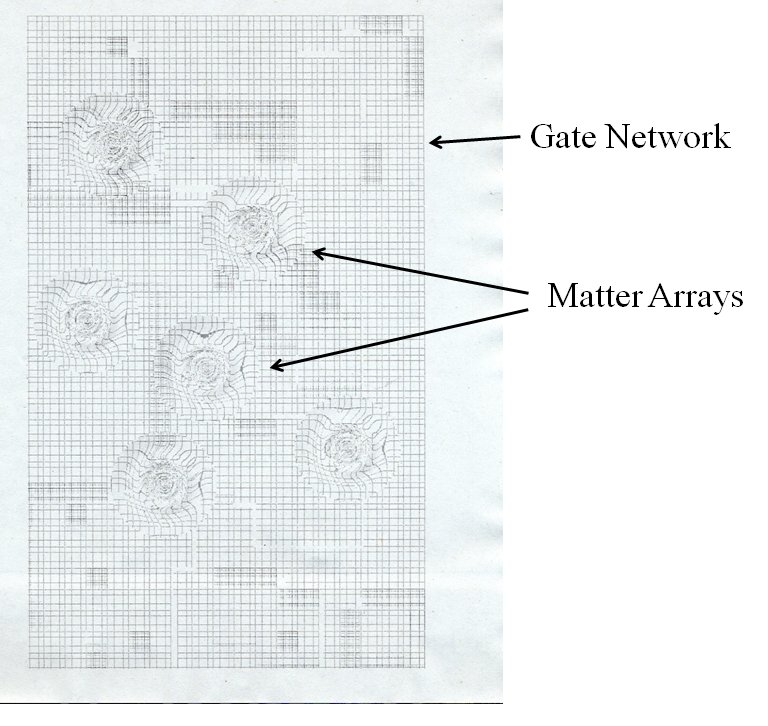}
\end{center}
\caption{Artist conception of two dimensional gate network showing how matter arrays can distort network where gate size is inversely proportional to energy.}
\label{fig:distort}
\end{figure}

It is useful to consider the Universe being only two dimensional like a sphere expanding into a third hyperspace as is often done to explain the Universe according to General Relativity as depicted in Figure \ref{fig:sphere}.
Picture then this sphere, (or another such closed object that does not necessarily have to be spherical), covered in a grid of lines that represent the gate network.
Where there is higher energy, be it from matter arrays possessing rest mass, or moving objects that have higher energy due to kinetic energy then their rest energy, the gate grid area shrinks according to $E = (\hbar c)/(\Delta x)$ where $\Delta x$ represents the 2D ``area''  of a gates associated with the matter array that is at rest or moving.
Figure \ref{fig:distort} depicts a portion of a two dimensional gate space network where distortion occurs due to the presence of energy or matter in an artist conception of the idea that is not being shown in a quantitatively accurate manner in this figure.
The conservation of energy then demands that as energy and matter shrink certain gate grid areas, other ones must expand that lose the energy somewhere on the sphere's surface as if the energy or matter is pulled on the grid.
Smaller grid areas on the sphere correspond to higher energy location where there is matter or energy quanta.
If the matter is moving then this pinched region of the network is also moving around the globe surface pulling on the other grids causing them to stretch and distort.
As such the grid on the sphere looks like a non-uniform grid of lines with a distorted mesh where there are spots where the mesh shrinks in the presence of matter and stretches apart in other places where there is less energy or in the wake of a moving object.
The distorted mesh represents lines of force between relativistic energy and mass.
The area of all gates together (volume in a three dimensional model) should be conserved according to $\Delta x = (\hbar c)/E$ since the energy is constant for the entire Universe where $\Delta x$ is here representing an area or volume in an abstract sense only. 
Different assumptions can be made about the grid on the sphere, or hypersphere in four dimensions, such as it is not rotating but that moving objects are simply distorting it as they move and different grid areas trade off volumes with each other as gates are transferred from one oscillator array (objects) to another through forces.
Or perhaps there is some rotational aspect of the grid itself that was imparted during the original expansion of the Universe, although this might be the same as assuming an angular momentum to the entire Universe that is something being debated by theorists with regards to predictions of General Relativity.
However, there may be a fundamental difference between matter and energy having a net rotation of some kind on a large scale in the entire Universe and the underlying gate space network through which matter and energy propagate if one prescribes to the authenticity of the gate space assumption.

Although it will not be proven here, it is reasonable to assume that within a matter array that behaves as an oscillator where all gates are quantum mechanically entangled with one another, that the total energy is equally shared between gates such that all gates have the same physical size and switching period $\Delta t$.
This might be necessary so that the gates ``fit'' together matching in energy and momentum within the array since $\Delta x$ is associated with momentum and $\Delta t$ is associated with energy conservation.
If two gates interact to swap information it is reasonable to assume that they must match in this way.
As photons interact with empty space, that could easily reside within the matter array itself in a multi-dimensional model, the empty space gate would also be influenced to match the size and period of the photon gate.
Photons travelling through space thus modify them distorting space and time as they propagate.
This could be seen as the quantum mechanical version of the classical field distortion associated with the presence of matter and energy in general relativity.

If gates must match in this way within a matter array oscillator, this would place quite a few restrictions on how matter can exist macroscopically.
It would be reasonable to assume that macroscopic objects would consist of many sub-matter arrays where the gates match only within each of these elementary arrays but where there is empty space between them so that they are not necessarily quantum mechanically entangled with one another outside of each elementary array.
Perhaps this is why we see atoms composed of only three types of particles, electrons, protons, and neutrons at this level of sub-atomic structure.
Even these particles are composed of more basic particles.
In this way Nature has arranged that gates into little elementary oscillators within oscillators so that only within an elementary oscillator do the gates match in size and period and are quantum mechanically entangled with one another.
Still, when a force is applied to such a collection of arrays that form a macroscopic object, photons can be transferred to all of the arrays fitting within them as required so that all of them transport themselves as per the Lorentz transformations each array acting in parallel with all of the others where quantum mechanical entanglement within each elementary array between all gates result in the motion as already derived for a single array.

Then we do not necessarily have all gates that make up a macroscopic object, even as small as an atom, being quantum mechanically entangled with one another, only the gates within each elementary array that form the elementary particles at the lowest hierarchy in the system.
One might conjecture that this level is the quark but perhaps even quarks have smaller arrays within them that have yet to be discovered.
It is clear, if the gate model is correct, that all matter must be quantum mechanically entangled between all gates at one level or another deep within the structure of matter as per the elementary photons and matter gates proposed in this mode.
Perhaps it has yet to be discovered experimentally, or perhaps elementary particles have already been identified that are constructed in this manner at the most fundamental level.
One might conclude that these gates and the elementary matter arrays they form, if they are indeed separate identifiable entities within elementary particles themselves, would form the most elementary entities in the Universe and other particles are merely collections of these elementary forms.
High energy particle accelerator experiments then may actually be arranging these elementary forms of gates and matter arrays in various ways at high energy that may not normally exist at lower energies in Nature.
At higher energies the gate sizes and time periods change compared to normal energy levels in everyday experience.
As such, in high energy experiments the rules for how the gates would merge to form matter oscillator arrays would alter compared to lower energy levels that normally exist in Nature.
As such particle accelerator experiments may be learning the rules as to how these gates join into matter and energy arrays (energy arrays become force exchange particles or mediators) but over a range of energies that might change dependent upon the energy of the interactions between the gates during collisions.
It does not necessarily mean that these particles or arrangements of gates into matter arrays found at higher energies in a particle accelerate would exist or take part in the elementary forces at lower energy levels.
It might be worthwhile designing experiments to reveal these more elementary forms to prove how they make up the particles we have discovered unless some of the particles we have already discovered are these elementary forms and to do so at energies we normally experience in Nature in everyday life.
Higher energy experiments would still be essential where data could be used to extrapolate to lower energies regarding how elementary photon and matter gates and arrays organize themselves in matter itself.

The ability to derive the central results of special relativity, that of the Lorentz transformations and the constancy of light propagation itself independent of a frame of inertial reference, appears to be a consequence of assuming that free space itself is composed of logic gates where energy quanta in the form of photons having the same gate structure as space itself can configure space itself to become those energy quanta or matter.
This provides a rather different view of the laws of motion.
Instead of thinking of matter and energy being entities in their own right, they become instead forms of empty space itself.
Movement is affected by the literal programming of empty space itself to become the energy or matter for brief periods of time over and over again as an object or energy propagates through space.
From this perspective, the particle or the wave is not the central focus of existence.
Existence is simply a connectivity property providing relationships between different kinds of information to effect its existence and configuration.
Space itself provides an interconnectivity network in which information can exist, move, or change interacting with other information content.
Matter and energy are quite literally empty space being no more solid than nothing itself.
The entire Universe is nothing at all but a huge lattice of circuit paths through which information can pass and interact.

When an object moves it is copied and destroyed over and over again as if being transported quantum mechanically from one place to the next identical to what has been achieved in laboratories where photons have been transported across labs over greater distances than the usual fundamental $\Delta x$ that already occurs naturally.
Between individual empty space gates what is taking place is not necessarily much different in concept to the so-called Science Fiction transporter apparatus where transport might take place over larger distances where the object is first copied into a new location and destroyed at the departure point.
What is clear from this model is that such a transporter beam might use empty space gates themselves to recreate the traveller at the new location without having to move anything but information itself.

Where this model may differ from other modern theories of space-time is the assumption that empty space itself is being configured to be matter and energy.
Additionally, the classical field results of special relativity are being derived directly from quantum mechanical considerations without any reference to the existing classical field theory of Einstein.
Existing relativistic quantum field theory (QFT), such as loop gravity, standard model and string theory, do not derive the classical forms of field theory, such as the Lorentz transformations, directly from quantum mechanical considerations, but rather assume the mathematical form as given by older classical theories of relativity then discretizing them forming the Lorentz symmetric group.
As such, other existing theories that attempt to combine relativity with quantum mechanics to formulate grand unified theories (GUT's), do so by combining classical and quantum mechanical results into a self consistent set of symmetrical groups representing laws of conservation.
However, it is well known that difficulties have arisen in attempts to unify all of the known forces of Nature into one form of physics.
The work in this paper does not attempt to do this, but rather attempts to show how the laws of motion can be obtained directly from the quantum mechanical properties by assuming that quantum logic gates or circuits are the fundamental information processing element of the Universe as opposed to particles, waves, loops, or strings etc.

Quantum gravity loop theory is probably the closest theory to the gate space theory presented here, but where the loops could easily be re-interpreted as being configurable Toffoli gates following the same spatial topology as loop theory.
Then all of the results obtained in loop gravity and standard model quantum field theory could be combined with the underlying principles presented here for gate space to provide a true unification of quantum mechanics and gravity.
It may not be necessary to completely reformulate everything in terms of gate space but just necessary to adopt the model presented here for the nature of space, energy and matter at its simplest most fundamental level.

Gate space then implicitly assumes that the entire Universe is one large Boolean logic function.
Sometimes this concept is referred to as the ``quantum world wide web'' in other theories that consider space to be made from logic circuits.
It is now mainstream accepted practise in quantum physics to assume that matter and energy are indeed kinds of quantum logic circuits where such circuits are now being exploited to make quantum computers.
However, it has not been clear just how one might derive the classical results of relativity from assuming such a model for space-time, the laws of motions, and gravity.
As such this paper attempts to repeat what Einstein did classically to introduce special relativity but using a purely quantum mechanical approach.
Hence, this paper could be seen as the quantum mechanical analog to Einstein's classical special relativity paper.

It is interesting that it is necessary to assume quantum mechanical entanglement, including the so-called ``spooky action at a distance'' that Einstein disliked so much in the theory of quantum mechanics, to derive his most famous theory, that of special relativity from purely quantum mechanical considerations.
Scientists, such as Einstein, had all of the necessary knowledge and experimental findings to derive the Lorentz transformations using the approach shown in this paper, but for some reason did not pursue it.
It is possible that reversible logic circuits were not in vogue at the time, however, it is not necessary to include these in the description of the derivation, just that photons mitigate information exchange between matter and empty space itself through quantum entanglement which is responsible for the inability of an object to reach the speed of light.
Perhaps the invariance of the speed of light as measured by Michelson and Morely that destroyed any possibility of an ether existing in the vacuum of space distracted scientists from considering that empty space itself might indeed have some properties that could be configured to become matter and energy without violating the light speed invariance property.

Gate space also implicitly assumes that the Universe has a no-moving background of gates through which all matter and energy quanta propagate themselves thereby enabling an independence of the speed of light from the velocity of its source or an observer.
This is not so strange when even in conventional models of modern physics, the invisible conceptual spatial co-ordinates of empty space itself was never considered to be moving either.
As such gate space is no different in this regard simply being another kind of background co-ordinate system but enhanced to include a natural interconnectivity as well as being configurable to become energy or matter.
In gate space all laws of motion are still relative to one another, however, the possession of energy is absolute which is not a relative entity in any existing theory of modern physics and which also is a necessary property to resolve the Twin Paradox problem in the theory of relativity.
In this paradox, the twin that has more inertial energy is the one that ages less.

\section{\label{sec:principles}Basic Principles\protect\\}

It is worthwhile reiterating several principles that were already introduced in this model of gate space, namely 1) the principle of non-simultaneity, 2) the principle of minimum existence, and 3) the principle of equivalent temperature.
The first principle was already described where nothing within a physical system can be said to take place at the quantum level simultaneously within the Heisenberg Uncertainty period $\Delta t$ depending upon the energy of the quanta involved and the energy resolution of the measurement.
Specifically this time was actually $\Delta t^2$ in the quantum entangled space-time continuum, however, we will refer to its square root informally to avoid confusion.
It is only necessary to consider that each interaction between gates, including with themselves, in a matter array or object that has matter and photon gates that trades information between gates including space gates, takes place sequentially with a characteristic time that is a function of the $\Delta t$ that is associated with the limits placed on observing or measuring a time interval by the Heisenberg Uncertainty Principle. 
This principle leads to the correct form of the Lorentz transformation that must necessarily be unique if the invariance of physics is to be maintained in all of physical laws.

The second principle of minimum existence simply states that energy and matter exist in their simplest form when there are two logic states available to the system.
In keeping with the concepts of using information theory to mirror physical laws, if the model presented for gate space in this paper is the simplest model that can explain the laws of physics, then it follows that there are real physical entities that correspond to this model that are at the most fundamental level within any system of matter or energy also being the smallest entities that can exist being indivisible.
In other words, there should be a quanta of energy that can be measured in an experiment that conforms to the model of a photon as described here where a real photon from the electro-magnetic force in three spatial dimensions may be more complex than a simple Toffoli gate possessing more complex behaviour and comprising several such gates.
Also, there should be a particle with rest mass in existence or that can be manufactured that conforms to the simplest oscillator that can exist in three spatial dimensions as a collection of Toffoli gates. 
Since there are three possible fundamental forms of entities within the Universe, force exchange particles we are referring to as elementary photons, mass not entangled with space, and space itself, the minimum model that adequately describes these three entities are reversible Toffoli logic gates.
Hence, all matter and energy can be modelled as hierarchical forms of such gates where the gates have properties of being observable or non-observable and being quantum entangled or not.
The properties of being observable or existing in reality is controlled by the control line $s_i$ logic value in the model circuit and quantum entanglement is the property of two gates exchanging control and input line logic values after the characteristic period of time $\Delta t$ has passed where the logic gates have cycled once through a logic state change.

The third principle of equivalent temperature is based on equations \ref{eq:En} and \ref{eq:E}.
This principle states that any physical system that is considered to be comprised of logic gates, can be modelled by any order $n$ of gates where the $T$ changes accordingly to keep the energy constant.
According to this equivalence principle, any physical system, including the entire Universe itself, can be reduced to its first order form where $n = 1$ with a maximum temperature $T$ as a means of analysis.
As the order $n$ changes, the temporal nature of the gate or gates changes accordingly such that all of the information in the system remains unchanged.
The trade-off between $n$, the total number of gates that have spatial extent, and $T$ to keep the energy constant amounts to a trade-off between the spatial and temporal composition of the information that is the system.
All physical systems can be reduced to two logic state single Toffoli gates with some maximum temperature T.
The time dependence of the gate may be quite complicated representing the system properties temporarily.

This last principle might provide some interesting possibilities if the gate space model is correct.
It might be possible to synthesize an object by modulating pure energy quanta with special temporal fluctuations that represent all of the information required to construct the object spatially.
Then these quanta would be allowed to cool adiabatically (i.e. keeping total relativistic energy constant) such that the temporal fluctuations become encoded spatially programming empty space itself to become matter arrays at some lower temperature where the order of the gates in the object $n$ is allowed to grow as temperature $T$ is allowed to lower keeping energy constant according to $E = nkT\ln{2}$. 
The particular mathematical forms of the original high temperature fluctuations would steer the nature and organization of the matter array composed of lower temperature matter and photon gates that compromise the final object.
Just what these temporal signals are might be discovered by heating and folding an actual object into it one or a few pure quanta in the reverse process.
This process might be referred to as space-time folding of an object where it is converted to perhaps a single vibrating energy quanta at high temperature where all of its information representing the object has become temporal in nature.
The energies and temperatures involved for large objects would be enormous and best left to some futuristic society.
In Sci-Fi this machine would be referred to as a ``replicator'' although it is more of a synthesizer or converter.
In some ways what is being done in nuclear reactors today where nuclear fuel turns into other elements through the fission process can be seen as a crude example of this concept, however, the process is hardly adiabatic.

\section{\label{sec:inflation}Inflationary Universe\protect\\}

At the so-called Big Bang, the entire Universe can be modelled as a single rapidly pulsating Toffoli gate with a complicated pattern of temporal behaviour switching back and forth between two logic states at a very high yet finite temperature.
Then for some reason not yet known, the Universe began to cool.
To keep the reaction adiabatic with constant energy, the original complex purely temporal properties of the Universe began to express themselves spatially as well as temporarily.
The minimum $\Delta t$ of any gate would have increased accordingly as energy was partitioned between an every increasing network of gates.
The energy itself would have become photon gates and mass gates with more and more empty space gates between as the Universe cooled.
Such a cooling process might have looked like an explosion to some imaginary outside observer, but in reality it was just simple cooling.

This cooling of the Universe is responsible for the so-called arrow of time providing its directionality.
In the beginning when the Universe was a single gate, which could be referred to as a singularity but that had a small non-zero spatial extent as a gate since the energy of the Universe is known to be finite and constant, its temporal signal would have contained both rational harmonic and irrational functional temporal components. 
However, since the number of gates of the Universe is always finite at any stage of its expansion, irrational functional information had to remain in purely temporal form, allowing only the rational functional information that can be represented exactly in finite form to distill out into spatial gate connectivity.
This irrational temporal information is responsible for the random noise one observes in temporal form in physical systems including thermal noise, flicker noise, etc.
Eventually it may be that all of the rational harmonic information may distill out into spatial gates leaving purely irrational information in temporal form.
At this point the Universe will reach its minimum temperature, its maximum gate number and its maximum expansion radius.
Also, only pure randomness will exist in temporal form maximizing entropy with vanishing ability for work to be done, the so-called Heat Death.

The continual creation of more and more space gates as the Universe cools results in the Hubble expansion that is observed astronomically.
It stands to reason that these gates are created more or less uniformly in space with a certain number of gates per distance.
As such one would expect the rate of expansion between any two points in the Universe due to this effect to increase with distance between the objects as is indeed observed.

\section{\label{sec:negative_energy}Negative Energy\protect\\}

Another phenomenon that results from the simple models presented in the previous section is that on average a gate takes up more spatial extent at lower energy for a given temperature.
As the Universe cools one could expect each existing gate plus new ones to experience an expansion.
This expansion can lead to a repulsive force that might explain the negative energy force believed to be responsible for objects receding faster than expected from predictions of some inflation models.

\section{\label{sec:action}Action at a Distance\protect\\}

In some ways quantum entanglement provides a near infinite dimensionality by every gate having the potential to be adjacent to each other gate through the ability to swap control line logic values, but not directly through the functional outputs that limit actual propagation of observable phenomena through space. 
For a quanta or object to actually move from one place to another it is necessary to do so via the interconnected output function $f_i$ and $\overline{f_i}$ connections in three dimensional space.
However, the ability to swap control line values with any gate no matter how far away through quantum entanglement gives every gate the non-local ability to instantly influence the possible outcomes of an action non-locally but not the ability to directly cause that outcome to occur instantaneously faster than the speed of light.

It is a subtle but important distinction between impacting the probability of an outcome denying certain ones faster than the speed of light, and actually causing it to occur in a measurable way.
If this model is correct, where the proper formulation of the Lorentz transformations depends upon it being so, then in principle it might be possible to effect faster than light travel by someone first travelling to a distant point not faster than light can travel but in the normal way.
Once there this first traveller would then do something at the intended destination that causes an outcome there to increase the probability of the empty space gates there to be configured to become identical to that of the next traveller's physical system at the destination instantly as in a kind of quantum transport.
If other travellers wished to travel it might be necessary to first send a sub-light signal to synchronize the wishes of the traveller, but the traveller themselves may not have to take the entire time themselves to end up at the destination once those at the other end knew of the traveller's intention to travel or their necessary co-ordinating information.
In this way causality may not be violated since it took the usual time for the event to take place but the traveller did not have to wait that long themselves to make the trip since their experience was instantaneous after the advance co-ordinating information arrived in the usual sub-light time.
Of course this is pure conjecture.

However, light speed transport of an object is not conjecture but should be possible if the information associated with the inputs to their Toffoli gates are transmitted in electro-magnetic form to configure empty space at the destination to become the object being transported.
The main point to understand here is that every possible object already exists everywhere in the Universe since everything is simply programmed empty space gates anyways.
As such only the pure programming information need be transmitted, that would correspond to quantum numbers as they are understood in conventional theories of quantum physics, the encodes the connectivity of the Toffoli gates of the original object turning on the control lines of the empty space gates used to reconstruct the object at the destination.

\section{\label{sec:dark_matter}Dark Matter\protect\\}

Another interesting possibility, if this model of space is accurate, that the control lines of an object might be turned off but the interconnections including the quantum mechanical entanglement between all of the gates comprising the object might be retained.
Such an object might still exist in empty space but would appear to have vanished.
This kind of object might be thought of as a kind of ``dark matter'' that cannot interact with photons in the same way as normal matter.
Having one of the input lines (either $s_i$ or $x_i$) in a logic state all of the time that prevents the other logic line from affecting the internal logic output state $f_i$ or $\overline{f}$ of each gate in the object prevents the object from being able to interact with normal matter or normal energy quanta in the same way.
However, it could be as complex as normal matter with regards to connections between gates forming oscillators within oscillators are per normal matter.
Also, it would contain energy, but a dark energy.
This relativistic dark energy would present mass and would also distort space thereby resulting in gravitational forces in a similar matter as normal energy and amtter composed of the same kind of logic gates but where the input logic lines do control the internal logic state.
Actually simply connected empty space itself could be thought of as having some mass but it does not distort space since it represents ``flat space'' compared to more interconnected object (normal matter and dark matter arrays) that do.
Just as in normal matter arrays, dark matter arrays might have the potential to interact with one another such that its matter gates decompose into photons since matter arrays may come about from photons of opposite momentum direction combining to become quantum mechanically entangled with one another.

\section{\label{sec:critical}Critical Behaviour\protect\\}

The gate space model may provide insight into the physics of critical behaviour such as phase changes between solids, liquids and gas and para magnetism.
Solids would involve matter array oscillators that are continuous within a solid, but if a critical number of photons are added in the form of heat then the entire oscillator would break up into smaller oscillators of the same size requiring the correct number of photons per separate oscillator within a liquid or gas to allow them to exist as separate interacting matter array separated by empty space gates alone.

We already know that matter has arranged itself into a hierarchy of oscillators one within the other in the form within electrons, protons, neutrons, within atoms within molecules within solids, liquids, gases, and plasmas to name a few or in the form of quarks, leptons, mediators, baryons, pseudoscalar and vector mesons.
Oscillators made from basic Toffoli gates connected in multi-dimensions have certain requirements to maintain oscillation in analogy to electronic oscillators formed from logic gates in rings.
The photons and matter gates formed from Toffoli gates discussed here should underpin all known particles by being combined in hierarchical oscillator forms.
It should be possible to relate energy and momentum conservation including all other laws of conservation that form the many symmetries of quantum field theory to combinations of gates forming such stable oscillator structures following the basic rules presented here including other rules normally associated with logic circuitry and oscillator theory but in higher ordered dimensional spaces as described to account for adjacency according to the laws of quantum entanglement.
The laws of physics derived in this paper simply form examples of what can be accomplished.
The success of assuming a logic gate structure of space itself lends credibility the approach. 
Instead of using group theory combined with quantum mechanics and field theory, perhaps Boolean algebra may be a better tool to develop a unified theory of everything.

\section{\label{sec:antimatter}Charge and Anti-Matter\protect\\}

Without going into details, it is worth point out that by making alterations to the basic Toffoli gate models of Figures \ref{fig:sgate} and \ref{fig:pgate} for instance by changing the input lines such that one uses PMOS and NMOS in different locations to invert either one or both of the $s_i$ control lines or the $x_i$ input lines it is possible to include more physics such as modelling charge and anti-matter.
Anti-matter could be modelled whereby the Toffoli gate control line $s_i$ is on for the opposite logic level as for normal matter as an example.
Then by their swapping control line information they would annihilate each other turning each other into photon gates provided the basic Toffoli gates that modelled photon and empty space gates were altered accordingly with interrupting the physics already developed here.

The point being made here without developing all of these ideas fully is that it is possible to find simple models to represent a rich physics to correspond with experiments.
The work presented in this paper proves that the basic laws of motion according to the laws of relativity can be modelled properly by using these concepts of a Toffoli gate representation of empty space.
Other observed experimental facts involving known interactions between the various particles and forces of the Standard Model can then be incorporated by using appropriate complexity by combining these gates as required where matter interacts with force exchange particles represented by gates configuring empty space gates through swapping of information from quantum mechanical entanglement as developed in this paper.
It is important to use the minimum complexity in gate structure to model the physics to arrive at the essential theories of Nature.

\section{\label{sec:artificial}Artificial Matter\protect\\}

Another interesting possibility worth conjecturing, provided the concepts of empty space being configurable into energy or matter quanta through quantum mechanical entanglement, is that it might be possible to synthesize artificial elementary particles from empty space gates that do not exist in naturally at the most fundamental level.
It is likely that actual space is compromised of more complex formations built up from the basic Toffoli gates presented including actual force exchange particles and matter arrays.
However, the simple structures assumed in this paper should be accessible through experiments that can then be used as basic building blocks to realize new kinds of force exchange particles and matter arrays by combining the basic Toffoli gates into more complex forms.
It may be possible to realize true artificial forms of matter in this way.
Hence, the basic Toffoli gates presented here would be sub-elementary particles even more basic then quarks and leptons.
It may be possible for instance that neutrinos are at the fundamental level being representable by a single or a few Toffoli matter gates which is why they have such low almost immeasurable mass.
The model of space gates presented here imply that all observable entities must have at least some mass and neutrinos have recently been found to have a small mass.

The model supports the notion that matter itself can be made from photons of opposite momenta colliding becoming quantum mechanically entangled with one another forming two matter gates in a matter array that loses its ability to engage empty space, or does so but only jiggles back and forth in a Brownian motion sense.
It should be possible to re-derive the Lorentz transformations by assuming that the matter gates $m$ are actually collections of such opposite momentum photons that can be treated as not being able to engage empty space in any particular direction but are still entangled with empty space because they are still really photons.
In this interpretation, photon gates and empty space gates are the only types of gates where matter gates are simply opposite directed photon gates entangled with one another into a local matter array oscillator.
Then matter itself is simply composed of such photons.

\section{\label{sec:stars}Stars and Black Holes\protect\\}

A star or sun is certainly a place with lots of photons and perhaps the photons themselves, as photon gates, not necessarily real electro-magnetic photons that may be more complicated combinations of Toffoli gates, fuse together to make the basic nuclei of matter. 
Stars are the programmers of Nature configuring empty space within them through photons making matter as well as emitting light.
If this is a correct interpretation then the Universe is a kind of Star Trek like hollow deck where photons cancel out to form matter.
This being the case it might be possible to make an artificial environment where carefully directed lasers emitting photon gates precisely aligned with opposite momentum to form solid matter only when the lasers are on that dissipate by being disentangled with one another through some kind of purposely introduced dissipative process.
It might be necessary to find ways to manipulate energy and matter at the ``gate'' level to accomplish this.

It is interesting to consider what might occur from a gate space perspective in black holes.
Matter arrays modelled using gates in three dimensions would normally have a lot of empty space gates within the array for normal matter densities.
In fact, most of the array would look like three dimensional ring oscillators within oscillators where most of the gates of the object would be empty space gates.
However, in a collapsing star, eventually one might envision, for a fixed amount of matter and therefore total energy and gate number, that gravity would compress the photon and matter gates to the point where they might completely fill in the empty space but still maintain their oscillator behaviour.
Eventually though there might be no free space left where the black hole is comprised of solid matter and photon gates.
It would then not be possible for the black hole to continue to collapse.
The only way to continue to collapse would be to increase in temperature reducing the number of gates according to $E = (m+l)kT\ln{2}$ where $m+l$ are the number of matter and photon gates.
If the temperature were to increase then eventually there would be only one extremely hot gate with a very high frequency.
There would be no photons to leave the black hole since it is only one gate and would be held in gate space by being smaller in size than the surrounding gates further away in empty space.
If it did not increase in temperature then its radius would stop shrinking and it is entirely possible to imagine that the photon gates would not propagate away if there are no empty space gates to use since matter gates cannot swap information with empty space and they are the only ones the photons have to swap information with.
Only photons are the edge of the black hole, or its event horizon, would have access to empty space.
Logic circuits are already used to model black holes so it is likely that the model presented here, if it is different to some extent, would also apply.
It is not clear, however, that these existing models derive the Lorentz transformations from pure quantum mechanical considerations or assume them first in classical form as other quantum gravity theories, which is an essential difference between this gate space model and other theories.
If other theories are not doing this then it might be worthwhile to incorporate the ideas in this paper with these other theories.
It should be possible to use the central ideas in this paper to modify any quantum gravity theory that has path integrals or logic circuit interactions to compute the increase path length or time delay associated with the Lorentz transformations from a quantum mechanical perspective.

\section{\label{sec:conclusions}Conclusions\protect\\}

In summary, it appears that the gate space model of physics may be useful to model a variety of physics being rich enough to take into account all that is known today.
It may be an inescapable conclusion that our Universe is indeed some kind of random digital simulation not unlike the giant Universe computer concepts of science fiction stories.
The fact that the model of empty space being configurable logic gates in itself that can be configured into photons or matter can so easily lead to fundamental laws of physics that pertain to both quantum mechanics and relativity using the same set of assumptions is in itself a compelling reason to perhaps accept this description as being more accurate than others.
Indeed, existing theories of physics seem to be making similar conclusions, albeit in more indirect ways, that might in themselves be modified to more clearly include a gate space concept.
Assuming a gate space model of the Universe where everything is made from what seems to have no substance seems to open up possibilities of the existence of other physical entities, beyond matter and energy that might also arise from configuring empty space.
It puts matter and energy that we normally associate with reality on a even footing with perhaps more ethereal entities such as empty space itself, or objects composed of empty space gates but that form matter arrays identical in form to real objects but where control line logic values prevent interaction with normal matter and energy in the same way.
Such objects may be dark matter, however, there may be other consequences associated with such objects that scientists would normally rule out as being possible but that most people naturally assume to exist, that being a spiritual world within our own.
One cannot rule out people's natural intelligence to sense the truth.
The Big Bang could be considered to have been first described in Genesis of the Bible where it states that the Universe began as ``Let there be light''.
It is also interesting that Genesis describes that creatures first began in the ocean and that plants preceded animals on land, both now considered facts by modern scientists.

The gate space is quite simple to understand, and many scientists might think too simple.
However, is it necessary for an accurate theory that describes the underlying principles of the Universe in a unified manner be complex, given that it appears to have come about on its own?
On the other hand, if the Universe can be best described as being a large interconnected set of logic circuits then it begs the question: who is the programmer?


\end{document}